\documentclass[twocolumn,numberedappendix]{openjournal}
\usepackage{amsmath,amsfonts}           

\usepackage{hyperref}

\usepackage{graphicx}       		
\usepackage{float}          		
\graphicspath{{././figures/}}  		

\def\lesssim{\mathrel{\hbox{\rlap{\hbox{\lower4pt\hbox{$\sim$}}}\hbox{$<$}}}}
\def\gtrsim{\mathrel{\hbox{\rlap{\hbox{\lower4pt\hbox{$\sim$}}}\hbox{$>$}}}}
\def\Myr{{\rm\,Myr}}
\def\Gyr{{\rm\,Gyr}}
\def\msun{{\rm\,M_\odot}}
\def\kpc{{\rm\,kpc}}
\def\pc{{\rm\,pc}}

\begin{document}


\shorttitle{The Impact of Feedback-driven Outflows on Bar Formation}
\shortauthors{Martin D. Weinberg}

\title{The Impact of Feedback-driven Outflows on Bar Formation}
\author{Martin D. Weinberg}
\affil{Astronomy Department, University of Massachusetts, Amherst, MA 01003}

\begin{abstract}
  We investigate the coupling between the temporal variation from
  central mass outflows and the bar instability. We show that
  gravitational fluctuations from AGN-driven mass outflows affect the
  radial motion of disk orbits.  The resulting incoherence in orbital
  phase leads to the disruption of the bar-forming dynamics.  Bar
  formation is suppressed in galaxies with repeated outflow events
  separated by \(10 < \tau < 200 \Myr\) with wind mass \(\sim 15\%\)
  of the disk within \(\sim 0.5\) scale lengths or 1.4\% of the total
  disk mass.  The work done by central AGN feedback is capable of
  reducing the amplitude or, with enough amplitude, destroying an
  existing bar.  Conversely, SF feedback that is distributed across
  the disk has insufficient spatial and temporal coherence to suppress
  bar formation.
  
  To model the dynamics of the coupling and interpret the results of
  the full N-body simulations, we introduce a generalization of the
  Hamiltonian mean-field (HMF) model, drawing inspiration from the
  Lynden-Bell (1979) mechanism for bar growth.  Our non-linear
  \emph{BarHMF} model is designed to reproduce linear perturbation
  theory in the low-amplitude limit. Without feedback, this model
  exhibits exponential growth whose rate depends on disk mass and
  reproduces the expected saturation of bar growth observed in N-body
  simulations.  We describe several promising applications of the
  BarHMF model beyond this study.
\end{abstract}


\section{Introduction}

\subsection{Background}

Bar are a ubiquitous feature of disk galaxies: at least 50\% of all
disk galaxies in the local universe are barred
\citep[e.g.][]{Aguerri.etal:2009}.  Bars were among the first
instabilities to be found and studied in N-body simulations
\citep{Hohl:1971,Ostriker.Peebles:73}.  They are predicted to be
important drivers of secular evolution
\citep{lynden72,Tremaine.Weinberg:84} and barred galaxies may
\emph{require} secular evolution to create their present-day
morphology \citep{Athanassoula:2003}.  In this way, their coupling to
the baryonic, and the dark-matter components of a galaxy through
cosmic time make their dynamical details an important diagnostic of
galaxy formation and evolution.  A significant fraction of the stellar
disk takes part in the bar instability, making bars morphologically
distinct and straightforwardly observed in the near-infrared.  Indeed,
recent analyses of JWST NIRCam images suggest that some bar systems
may have been active for many gigayears \citep{Guo_2023}.  An
understanding of the formation time and duty cycle of bar-driven
evolution requires a thorough understanding of the dynamical
mechanisms that affect bar evolution.

Both by technical necessity and experimental design, most theoretical
and numerical studies of barred galaxy dynamics begin with quiescent
equilibria chosen to be unstable to bar formation.  Some of the
dynamical principles governing bar formation and subsequent secular
evolution are well-understood after decades of study
\citep[see][]{Binney.Tremaine:2008}.  While our dynamical insight
follows from such carefully controlled dynamical experiments
\citep[e.g.][]{Petersen.etal:2019, Petersen.etal:2021}, we are
well-aware that galaxies continue to accrete and lose mass through
cosmic time, but we have not given this time dependence detailed
attention dynamically.  In particular, a wide variety of
time-dependent feedback processes are necessary to produce the
galaxies in cosmological simulations
\citep[e.g.][]{Ceverino.Klypin:2009, Kormendy.Ho:2013}.  For example,
the work done on the gravitational potential by this feedback has a
natural time scale and amplitude.  The time scale is approximately
10--200 Myr based on theory and estimated from observations
\citep{Heckman.etal:1990,Hopkins:2010}.  The amount of mass lost from
the disk, and hence work done on the disk gravitational potential is
much harder to estimate.  Star formation efficiencies in a star
forming event are estimated to be \({\cal O}(0.1)\) and estimates of
mass outflow rates from AGN feedback range from 1 to 5 times the star
formation rate \citep{Combes.etal:2013,Combes:17}.

The bar instability itself results from the precession of apocenter
positions of nearby orbits toward its self-induced quadrupole
distortion.  If a non-trivial fraction of the remaining gas is heated
or launched into halo during an AGN outburst or strong star-formation
event, the work done on the gravitational potential changes the
orbital actions of stars in the disk.  These event time scales are
similar to the characteristic orbital times in the disk.  Thus, the
resulting mass-outflow-driven fluctuations can affect the orbital
coherence required for bar formation.  We expect this coupling to be
particularly important during at redshifts of \(z = 2\)--\(3\) often
called \emph{cosmic noon}.  During this rapid baryonic accretion
phase, galaxies formed about half of their current stellar mass
\citep{Madau.Dickinson:2014}, gas advection will be strong efficiently
feeding central super-massive black holes (SMBHs), and the effect of
feedback will be maximal and most likely significant source of
gravitational fluctuations.  While widely explored in suites of
galaxy-formation simulations, the details of multi-scale physical
processes are not well determined.  An overall, general description of
the coupling between the feedback fluctuations and the bar instability
mechanism and estimating the importance of feedback to bar formation
is the main goal of this paper.

\subsection{Feedback processes}

There are two main feedback mechanisms in cosmological simulations
that have regulatory affect on the baryonic history: active galactic
nuclei (AGN) and star formation (SF). Presumably, both play a role in
the self regulation necessary to achieve the observed characteristics
of present day galaxies.  Both AGN winds and the interaction of
supernova feedback with their surrounding gas through the injection of
energy and momentum can lead to a feedback loop regulating ongoing
evolution and outflow.  An effective feedback loop requires removing a
large fraction of the disk's cold gas by winds, heating and fountains.
A variety of sub-resolution prescriptions have been employed to
achieve this.  Regardless of the details of the microphysics, these
outflow events are episodic, not continuous.  Because these events
tend to be centrally dominated with a cadence of
\(10 < \tau < 200\,\mbox{Myr}\), outflow events will provide
gravitational fluctuations on dynamical timescales to the axisymmetric
field of the disk.  Similar to the regulation by feedback, the mass in
the outflow need not escape from the galaxy to have affect on bar
dynamics since most of the gravitational work occurs over the first
kiloparsec.

Feedback processes and galactic winds specifically have been
well-reviewed over many years \citep[e.g.][]{Veilleux.etal:2005,
  King.Pounds:2015, Zhang:2018}.  The importance of the physical
details in modeling SF regulation is also well appreciated.  For
example, \citet{Governato.etal:2007} remarked that subgrid star
formation feedback tended to make the disk more stable against bar
formation by reducing the mass of the disk. Since then, simulations
with higher resolution and sophisticated subgrid prescriptions have
demonstrated large variance and intricate coupling between outflow,
re-accretion of gas, and subsequent star formation with bar formation,
even in the same simulation suite \citet[e.g.][]{Zana.etal:2019}.

\subsection{Coupling to disk bars and caveats}

The goal of this paper is a characterization of the coupling between
feedback-induced fluctuations in the gravitational field and the
galactic dynamics of bar formation.  The proposed model is very simple
and crude by today's standards for \emph{subgrid} physics in
cosmological simulations.  It is a simple Markov jump process for
central feedback-driven outflow. This models feedback as a sudden loss
of mass at Poisson-process intervals with an exponential replenishment
time.  Thus, the disk mass remains constant on average without
explicitly following the mass settling or reaccretion itself.  We then
investigate the fluctuation amplitude and time scales parametrized by
this simple model that are required to disrupt bar formation.  We
choose an exponential disk and live dark-matter halo with masses
typical of typical of a Milky-Way-like galaxy at \emph{cosmic noon}
for our simulations.

The background model adopted for this study is not intended to be
realistic in the cosmological structure formation sense; neither our
disk or halo accretes new material in time.  For this reason, we focus
on evolution times \(<4\,\mbox{Gyr}\).  Similarly, our simulations do
not include satellite and sub-structure interactions.  We emphasize
that our feedback process is no more than a \emph{fitting formula}; it
does not consider the physical mass-loss time scale or the impact of
the feedback on the structure or populations in the disk.  Rather,
this suite is intended explore the dynamics of the coupling mechanism
and to address two particular questions: (1) What is the stochastic
frequency required to disrupt bar formation?  (2) What is the
amplitude of the work done required to disrupt bar formation?  While
inspired by feedback, the proposed dynamical mechanism is generic and
intended to provide a framework for interpreting more realistic
simulations.  The mechanism proposed here has thematic overlap with
the suggestion by \citet{Ogiya.Mori:14} that resonant coupling between
a feedback-excited density wave can promote the cusp-core transition.
The main dynamical difference is we couple the gravitational field
fluctuation from a stochastic outflow directly to the dynamics of bar
orbits.

Our \emph{primary} galaxy model forms a bar in the absence of
potential fluctuations with a disk mass of approximately
\(M_{disk} = 5\times10^9\msun\) with \(M_{halo} = 10^{12}\msun\).  The
system is not strongly bar unstable owing to its relatively low,
\emph{sub-maximal} disk mass.  We intentionally choose a bulge-free
model with a rising rotation curve inside of 6 kpc to promote bar
formation within a few gigayears. Both growth rates and subsequent
pattern speed changes will are model dependent.  We show that bar
formation is suppressed for strong central outflows with
characteristic periods in the range \(10 < \tau < 200\,\mbox{Myr}\).
Sufficient \emph{blow out} amplitude is approximately \(15\%\) of the
baryonic mass inside of \(0.5\) disk scale lengths.  Our super-bubble
model \citep{Kipper.etal:25} \emph{does not} suppress bar formation
for parameters typical of starburst rates.  This suggests that
large-scale central outflows are necessary for the dynamical model to
be effective.

\subsection{A theoretical bar model}

To provide dynamical insight for the feedback coupling to bar
dynamics, we present a new hybrid perturbation-theory-based
simulation.  Specifically, we adopt the tools of Hamiltonian
perturbation theory to isolate the single degree of freedom that
causes the outer apsides of disk orbits to collect along a diameter.
This is the bar-formation mechanism described in
\citet{Lynden-Bell:1979}.  This isolation is derived by performing a
change of angular coordinates where one of the angles describes the
motion with respect to the inner-Lindblad resonance.  Near resonance,
the angular speed of this angle is much slower than the rapid
variation of the star about its orbit radially.  This allows us to
average of the radial motion and reduce the dimensionality of the
equations of motion.  Appendix \ref{sec:Hpert} develops this idea into
a particle simulation method.  Although based on linear perturbation
theory, the coupling is nonlinear and related to the Hamiltonian Mean
Field model \citep[e.g.][]{Chavanis.etal:2005}.  This allows
investigation of both the linear and nonlinear phases of bar growth.
We use this model in Section \ref{sec:barhmf} to verify the dynamical
interaction between feedback and bar formation.

\subsection{Plan}

The plan for this paper is as follows.  Section \ref{sec:methods}
describes our galaxy mass profiles (Sec. \ref{sec:model}), summarizes
our N-body simulations of the disk and halo (Sec. \ref{sec:exp}),
defines the parametrization of the feedback mechanisms for both
central AGN-type and spatially distributed SF-type outflows
(Sec. \ref{sec:pert}), and the describes the new hybrid
perturbation-theory based simulation (Sec. \ref{sec:pertsim}).  The
results are described in Section \ref{sec:results}, beginning
with a demonstration of the main result in Section \ref{sec:SFmain}
followed description of the primary model without feedback in Section
\ref{sec:primary}.  Sections \ref{sec:eps} and \ref{sec:tau} describe
the amplitudes and frequencies of the feedback process necessary to
suppress the bar instability.  Section \ref{sec:barhmf} presents a
dynamical mechanism for the noise coupling in the reduced model and
compares it to the results from the full N-body simulations. We
conclude with a summary and discussion in Section \ref{sec:conclu}.

\newcommand{\EXP}{{\sc{exp}}}

\section{Methods}
\label{sec:methods}

\subsection{Galaxy model}
\label{sec:model}

We use \emph{virial} units for length and mass, \(R_{vir}\) and
\(M_{vir}\) respectively, with gravitational constant \(G=1\) for all
simulations described here and refer to those as \emph{system units}.
We present our results scaled to the Milky Way throughout to give
physical context.  We adopt \(R_{vir}=300\kpc\) and
\(M_{vir}=1.4\times10^{12}\msun\) and use gigayears (Gyr) for time
units.  The system time units are
\(T_{vir}\equiv(R_{vir}^3/GM_{vir})^{1/2} \approx 2\) Gyr; or 1 Gyr is
approximately 0.5 system time units.

\subsubsection{Dark-matter halo}
\label{sec:halo}

For our primary model, we adopt a modified NFW \citep{Navarro.etal:97}
dark matter halo with concentration \(c\), whose density is given by
\begin{equation}
  \rho_{\rm halo}(r) \propto \left(r+r_c\right)^{-1}\left(r+r_s\right)^{-2}
\label{eq:nfw}
\end{equation}
where \(r_s\equiv R_{\rm vir}/c\) is the scale radius, and
\(r_c\ll r_s\) is a radius that sets the size of the core. In this
paper, the core radius is numerical convenience only and set to a
value smaller than any astronomically relevant scale, typically
several parsecs.  We adopt \(c=15\) for our dark matter halo which
provides an acceptable representation of the rotation curve when
scaled to the Milky Way.  This is slightly more concentrated that the
mean from \citet{Diemer:2015} but within the variance measured from
simulations.

In practice, we choose a more general profile family that includes the
NFW model to explore the dynamical effect of changing the background
profile from shallow to steep and enforce a finite total mass.
Specifically, we modify a two-power halo model with an error function
truncation of the following form:
\begin{equation}
  \rho_{\alpha\beta}(r) = \frac{\rho_0 r_s^{\alpha+\beta}}{
    (r+r_c)^{\alpha}(r+r_s)^{\beta}}
  \left\{\frac{1 + \mathrm{erf}\left[ (r_t - r)/r_w\right]}{2}\right\}.
\label{eq:twopower}
\end{equation}
where \(\rho_0\) is a normalization set by the chosen mass.  This is
equivalent to equation (\ref{eq:nfw}) for \(\alpha=1\) and
\(\beta=2\), and \(r_t=\infty\).  The outer truncation radius \(r_t\)
and truncation width \(r_w\) are chosen to be 300\kpc\ and 60\kpc,
respectively.  For our primary model, we set \(\alpha=1\),
\(\beta=2\), \(r_c=4.5\pc\).  We explore the effect of different
\(\alpha\) in Appendix \ref{sec:hmftests}.

While one may debate the merits of a purely collisionless dark matter
particle, axion dark matter, `fuzzy' dark matter
\citep{FDM:2000,Hui.etal:2017}, or other descriptions of dark matter,
the necessity of exploring the interaction between baryonic and unseen
components via gravity may not be ignored.  Dark matter will respond
gravitationally to baryonic matter and different dark-matter responses
can produce observable signatures.  Rigidity in halos has been noted
to limit bar growth \citep{polyachenko16,sellwood16} as is clear from
dynamics of secular evolution \citep{Athanassoula:2003}, and so we
emphasize that the live halo inclusion is crucial to recover the
dynamics of the standard \(\Lambda\)CDM scenario.  Specifically, both
the fraction of gravitational support from the halo and the resonant
coupling will affect bar growth, although we will not explore that
sensitivity in this paper.  There are many other parameters that may
be adjusted in the creation of a halo which we do not explore here:
triaxiality and spin are two that are addressed by others
\citep{athanassoula13,aumer16a,collier18,collier19}. However, we do
not expect either of these to qualitatively change the dynamical
features explored in this work.

\subsubsection{Disk}
\label{sec:disk}

The simulations begin with an exponential disk of surface density
\begin{equation}
  \Sigma_{\rm disk}(r) = \frac{M_{\rm d}}{2\pi a^2}
\exp\left(-\frac{r}{a}\right)
\label{eq:exponentialdisk}
\end{equation}
where \(M_d\) is the disk mass, and \(a\equiv0.01R_{\rm vir}=3\kpc\)
is the disk scale length.  We assume that the disk remains thin with
no vertical response to further restrict the dynamical degrees of
freedom for this simple proof-of-concept investigation.  This is often
called the \emph{razor-thin} limit.  We have repeated a small set of
key simulations described in Section \ref{sec:AGNmain} with a full
three-dimensional disk and the results are nearly identical, as
expected.

We select the initial positions in the disk via acceptance--rejection
algorithm using equation (\ref{eq:exponentialdisk}). We select the
velocities by solving the Jeans equations \citep{Binney.Tremaine:2008}
with an axisymmetric velocity ellipsoid in the disk plane
(\(\sigma_r\equiv\sigma_\phi\)). We characterize the radial velocity
dispersion using the Toomre \(Q\) parameter,
\begin{equation}
\sigma_r^2(r) = \frac{3.36\Sigma(r)Q}{\Omega_r(r)}
\label{eq:toomre}
\end{equation}
where \(\Sigma(r)\) is the disk surface density, and the radial frequency, \(\Omega_r\), is given by
\begin{equation}
\Omega_r^2(r) = r\frac{d\Omega_\phi^2}{dr}+4\Omega_\phi^2.
\end{equation}
where \(\Omega_\phi\) is the azimuthal frequency.  See Section
\ref{sec:IC} for additional details.  The circular velocity curve for
this model is shown in Figure \ref{fig:rotcurve}.  The rise in the
inner few kiloparsecs promotes bar growth by design.

\subsection{N-body method}
\label{sec:exp}

The BFE method for N-body simulation uses a basis of
\emph{biorthogonal function} pairs constructed from the eigenfunctions
of the Laplacian. The pair solves the Poisson equation,
\(\nabla^2\Phi_j = 4\pi G\rho_j\), with the inner product
\((\Phi_j, \rho_k) \propto\delta_{jk}\).  Projecting the density
distribution implied by an ensemble of point-positions onto the basis
yields the gravitational potential.  Forces immediately follow.  This
method was introduced by \citep{Clutton-Brock:72}, inspired by the
mathematical tools used for the joint analytic solution of the
collisionless Boltzmann equation and the Poisson equation.
\citep{Clutton-Brock:72, Clutton-Brock:73, Hernquist.Ostriker:92} all
derive recursion relations for biorthogonal bases using various
transformations and sets of special functions.  This method has the
practical advantage of automatically filtering variations in the field
to the largest spatial scales.  We choose a fixed number of basis
functions to represent structural features on characteristic scales
much larger than the inter-particle spacing such as arms and bars in
the disk and dark-matter wakes in the halo.  The resulting algorithm
has computational complexity \(\mathcal{O}(N)\); that is,
gravitational field evaluation scales linearly with particle number.
The main disadvantage of this approach is that it lacks general
flexibility and depends on an expansion center.  Nonetheless, it is an
excellent choice for studying the evolution of systems to modest
perturbations.

\citet{Weinberg:99} proposed a numerical solution of the underlying
Sturm-Liouville equation to construct eigenfunctions whose
lowest-order basis function pair matches the desired background
equilibrium system. This extends the flexibility of the BFE approach
by widening the suitable family of equilibrium models to cover cases
that did not match the available recursion-relation bases.  The full
method is described precisely in \citet{Petersen.etal:2022}.  This
approach along with a suite of analysis tools tailored to BFE
simulations is implemented in \EXP\ \citep{EXP}, our publicly
available BFE N-body and analysis code.\footnote{The \EXP\ code suite
  is publicly available on GitHub (see
  \url{https://github.com/EXP-code/EXP.git}) and full documentation is
  available from ReadTheDocs (see
  \url{https://expdocs.readthedocs.io}).}  We implement bases for a
variety of geometries, including spherical, two- and
three-dimensional cylindrical, slabs, and rectangular boxes.

For the spherical case, \EXP\ optimally represents the BFE with radial
basis functions determined by the target density profile and spherical
harmonics. The lowest-order \(l=0, m=0\) basis function matches the
potential and density of the equilibrium.  The overall fields of the
halo are described by
\(\left(l_{\rm halo}+1\right)^2\times n_{\rm halo}\) terms, where
\(l_{\rm halo}\) is the maximum order of spherical harmonics and
\(n_{\rm halo}\) is the maximum order of radial terms per \(l\) order.
For simulations and analyses in this paper, we use a maximum harmonic
order \(l_{\mbox{\tiny max}}=6\) and maximum radial order
\(n_{\mbox{\tiny max}}=20\) using the Sturm-Liouville basis
conditioned on each particular input model.  While the quality of the
expansion is center dependent, the basis can follow any displacement
from the center for sufficiently large particle number and large
values of \(l_{\mbox{\tiny max}}\) and \(n_{\mbox{\tiny max}}\).  The
choices of \(l_{\mbox{\tiny max}}=6\) and \(n_{\mbox{\tiny max}}=20\)
allow us follow the small center displacements seen in our
simulations.  A discussion of convergence for the NFW halo used here
is found in \citet[Section 3]{Petersen.etal:2022}.  The resulting
series of coefficients fully describe time dependence of the
gravitational field produced by the N-body evolution\footnote{This
  method will work for triaxial halos as well.  For triaxial halos,
  the target density can be chosen to be a close fitting spherical
  approximation of the triaxial model.  The equilibrium will require
  non-axisymmetric terms but the series will converge quickly.}.

Appropriate basis functions for a three-dimensional cylindrical disk
are described in \citet[Section 3]{Petersen.etal:2022} along with a
study of convergence properties.  Here, we use a recently-implemented
special-purpose two-dimensional cylindrical disk basis that is
described in Appendix \ref{sec:2ddisk}. The method and performance
well described by the \citet{Petersen.etal:2022} study restricted to
\(z=0\). This focuses our dynamical attention on the two degrees of
freedom essential to create a bar: the radial and azimuthal motion. We
have demonstrated that same results are obtained for the full
three-dimensional simulations in selected cases.  Future work will
include coupling to the vertical motion from possibly asymmetric
outflows.

In summary, \EXP\ allows for a straightforward calculation of the
gravitational potential from the mass distribution through time.  The
key limitation of the BFE method lies in the loss of flexibility owing
to the truncation of the expansion series: large deviations from the
equilibrium disk or halo will not be well represented. Although the
basis is formally complete, our truncated version limits the
variations that can be accurately reconstructed. Despite this, basis
functions can be a powerful tool to gain physical insight; analogous
to traditional Fourier analysis, a BFE identifies spatial scales and
locations responsible for the model evolution.  In particular, this
paper investigates relatively low-amplitude, large-scale distortions
which are well represented by the expansion.

\subsection{Initial conditions}
\label{sec:IC}

We generate realizations with \(10^6\) disk particles and \(10^7\)
halo particles from the models in Sections \ref{sec:halo} and
\ref{sec:disk} using the following algorithm:
\begin{enumerate}
\item The halo gravitational potential is modified to contain the
  monopole component of the exponential disk.  The halo phase space
  distribution is then realized using Eddington inversion
  \citep[e.g.][]{Binney.Tremaine:2008}. This helps improves the
  initial equilibrium.
\item The disk phase space is generated using Jeans' equations as
  described in Section \ref{sec:disk}.  Gaussian velocity selection
  rather than velocity selection from the exact phase-space
  distribution function will introduce disequilibrium to start.  This
  is mitigated by the relaxation in Step \ref{step:diskgen}
  described below.
\item The halo phase-space distribution is \emph{relaxed} in the
  presence of the potential generated by the disk particle
  distribution using the \EXP\ potential solver for 4 Gyr (\(2
  T_{vir}\)).  The force felt by the halo from the halo's own self
  gravity is restricted to \(m=0\) terms during this relaxation phase.
  The force felt by the halo from the disk is the disk's gravity at
  \(T=0\) for the axisymmetric component \(m=0\) only.
  \label{step:halogen}
\item The disk phase space is regenerated using Jeans' equations as
  described in Section \ref{sec:disk} using the gravitational
  potential of the halo distribution at the end of the relaxation
  step.
\item The disk phase-space distribution is then relaxed in the
  presence of the potential generated by the halo particle
  distribution using the \EXP\ potential solver for 4 Gyr.  The force
  felt by the disk from the disk's own self gravity is restricted to
  the \(m=0\) term.  The force felt by the disk from the halo is the
  halo's gravity for the \(m=0\) axisymmetric component only frozen at
  the beginning of this simulation.
  \label{step:diskgen}
\item The phase space distributions for the halo at the end of Step
  \ref{step:halogen} and the disk at the end of Step
  \ref{step:diskgen} are the phase-space initial conditions for the
  simulations reported here.
\end{enumerate}
This algorithm is enabled by the basis-function approach that
underlies \EXP.  In particular, the BFE approach used in
\EXP\ provides a separate basis for each component (disk and halo,
here) and each the expansion terms for basis may be applied
selectively or frozen in time.  For example, in Step
\ref{step:halogen} we allow only the axisymmetric force of the halo to
obtain the new \emph{spheroidal} equilibrium in the presence of the
disk while preventing the natural halo modes from influencing the
initial conditions.  The resulting initial conditions are very close
to feature-free to start.  The initial condition realization described
here has also been discussed in \citet{Petersen.etal:2022}.

\subsection{Perturbation scheme}
\label{sec:pert}

We describe each of our two stochastic outflow models below.  In
Section \ref{sec:central}, we simulate a central outflow, applicable
to both an AGN or nuclear superwind \citep{Romero:18}, by modulating
the mass of the stars according to a specific schedule that is tracked
by the particles themselves.  This could equally well have been
implemented with negative mass disk-shaped particles that projected on
the expansion coefficients directly.  We implemented this coefficient
projection scheme as a check on particle-mass adjustment algorithm and
obtained similar results.  To simulate the typical SF outflow process
in a more realistic way, we implement diffuse gas bubbles with the
spatial distribution of a snowplow-phase mass shell (Section
\ref{sec:super}).  A suite of simulations suggests that these super
bubbles have very little effect on bar formation for
astronomically-relevant parameters, suggesting that the central
fluctuation of the global gravitational potential is a necessary for
effective coupling to the bar instability.

\subsubsection{The central outflow model (AGN)}
\label{sec:central}

We assume that an AGN or nuclear starburst can be modeled as a
stochastic process in the inner galaxy disk that causes a sudden loss
of fraction \(\epsilon\) of the enclosed mass within radius
\(R_{burst}\equiv\lambda a\) where \(a\) is the disk scale length. In
other words, mass from the outflow is lost from an inner disk with radius
\(\lambda a\). Experiments performed as part of Section
\ref{sec:results} show that the dynamical effect of the mass loss is
only weakly dependent on \(\lambda\) for \(\lambda\lesssim0.5\). The
lost mass is assumed to be fall back onto the disk over some
characteristic period \(P_{acc}\).  This emulates a radiatively driven
outflow and the subsequent resettling of cooling gas.  The
perturbation is modeled mathematically as a Markov jump process with
two characteristic frequencies: one that describes the outburst
frequency, \(1/\tau_1\), and one that describes the settling time,
\(\tau_2=P_{acc}\).  Let \(\tau_{dyn}\) be the dynamic time
scale.  Physical consistency implies that
\(\tau_{dyn}\lesssim\tau_2\).

The algorithm is implemented in \EXP\ as follows:
\begin{enumerate}
\item Let \(t\) be the current simulation time and let \(t_\ast\) be
  the time at the last outflow event.  At every time step, an outflow
  event occurs with exponential probability (Poisson)
  \begin{equation}
    P(t)=e^{-(t-t_\ast)/\tau_1}.
    \label{eq:Pevent}
  \end{equation}
  This implies that the likelihood of a new event at the current time
  \(t\) is \(P(t>t_\ast) = 1 - e^{-(t - t_\ast)/\tau_1}\).  Therefore,
  we may randomly generate the time for the next event as
  \(t_{+} = t_\ast -\tau_1\log({\cal R})\) where \({\cal R}\) is a
  random variate in \((0, 1]\).
\item An event occurs when \(t\ge t_{+}\).  At that time, all star
  particles within radius \(\lambda a\) lose a fraction of their mass,
  \(\epsilon\).  We set the particle event time
  \(t_{+}\rightarrow t_\ast\).  Each star particle carries the time of
  its last event.
\item Each star gains back mass according to:
  \begin{equation} m =
    m_o\left[1 - \epsilon e^{-(t-t_\ast)/\tau_2}\right],
    \label{eq:Mdecay}
  \end{equation}
  where \(m_o\) is the mass of the star at \(t=0\).
\item At the start, we choose \(t_{+} = -\tau_1\log({\cal R})\) and
  \(t_\ast=-\infty\) for all particles.  This implies that the mass
  per star particle ranges between \((1-\epsilon)m_o\) and \(m_o\) and
  is \(m_o\) to start.  We emphasize that changing the mass per
  particle is a device to adjust the gravitational profile of the
  inner galaxy.  The individual trajectories only change through this
  mass adjustment to the mean gravitational field.
\item The AGN or central event frequency, \(1/\tau_1\) and the mass
  replenishment time, \(\tau_2\), may be different.  For results
  reported here, we assume that \(\tau_2 = \tau_1/2\) so that the disk
  recovers most of its mass on average before the next AGN
  event\footnote{Our model demands \(\tau_2<\tau_1\) to represent a
    fluctuating quantity; for \(\tau_2\gg\tau_1\),
    \(m\rightarrow m_o(1-\epsilon)\) and remains nearly constant.
    However, if \(\tau_2\) is much smaller than a orbital time, the
    mass resettles before stellar trajectories can be affected by the
    change in gravity. This time scale is limited from below by the
    outflow wind speed, which is typically no more than five times the
    typical circular velocity.  Similarly, \(\tau_1\) is typically
    less than an orbital time.  This implies that
    \(\tau_{dyn}\lesssim\tau_2\lesssim\tau_1\).}.  Tests show that the
  end results are very weakly dependent on the precise value of
  \(\tau_2\) for \(\tau_2={\cal O}(\tau_1)\).
\end{enumerate}
To reiterate, this algorithm conserves mass over long time intervals;
we assume that some process exists to return the \emph{lost mass} to
disk for \(\tau\gg\tau_2\).  This may through accretion of hot halo
gas or the settling of a galactic fountain gas. In any case, the mass
loss rate
\[
\dot{M}_{loss} = M_d(\lambda a)\epsilon/\tau_1,
\]
where
\[
M_d(R) = M_d\left[1 - (1+R/a)e^{-R/a}\right],
\]
may be larger than the typical mass-loss rate measured in galaxy
winds.  The gravitational fluctuation resulting from this mass loss
scales as \(G M_d\epsilon/a\).  Any variation in \(\lambda\) is
equivalent to a fixed value of \(\lambda\) and a commensurate
variation in \(\epsilon\).  Therefore, we assume a fiducial value,
\(\lambda=2/3\), for many of the runs considered here with
\(M_d=5\times10^9\msun\).  This implies that the mass affected by
feedback is
\begin{equation}
  M_d(2a/3) = 0.144 M_d \epsilon
  \label{eq:deltaM}
\end{equation}
and the mass loss occurs in the inner \(2\kpc\) of the galaxy.  As an
astronomical parameter, this radius is unrealistically large for an
actual AGN-driven outflow.  However, the axisymmetric (\(m=0\))
component of force felt by the trajectories that comprise the bar are
dominated by the lowest order terms in the \(m=0\) expansion.
Decreasing \(\lambda\) tends to increase the change in the higher
order terms and but not affect the change from mass loss in the
lower-order terms.  This explains the measured insensitivity to
smaller values of \(\lambda\).  A larger value of \(\lambda\) allows
the change in the gravitational field to be well-resolved by
the basis function expansion.

As an internal check, we also include the work done by an outflow wind
by replacing Steps (2)-(4) in the algorithm by a direct modification
of the gravitational field expansion coefficients as follows.  When an
event occurs at time \(t_j\), we create a negative mass disk with
radius \(\lambda a\) and surface density
\begin{equation}
  \Sigma_\epsilon(t) = -\epsilon \frac{M_d(\lambda a)}{\pi(\lambda a)^2}
  e^{-(t-t_\ast)/\tau_2}.
\end{equation}
The BFE coefficients describing this mass outflow are
\begin{equation}
\delta a^{\epsilon}_k(t) = 2\pi\Sigma_\epsilon(t)\int_0^{\lambda a}
dr r \Phi^\ast_{0k}(r)
\end{equation}
following the development in Appendix \ref{sec:2ddisk}.  These
\(\delta\mathbf{a}^\epsilon(t)\) values are added to the field
expansion coefficients from the particles in the simulation. The value
of \(t_\ast\) is updated whenever an event occurs.  We repeated the
simulations for a number of cases described in Section
\ref{sec:AGNmain} with no significant changes to the bar formation
trends.

\subsubsection{The super-bubble model (SF)}
\label{sec:super}

The mechanism from Section \ref{sec:central} describes a stochastic
loss of mass that does work on the center of a disk galaxy.  We
consider an alternative model inspired by the SF process specifically:
super-bubble creation by strong SF events scattered throughout the
galaxy disk.  We model super-bubbles as spherical shells of gas
describing the late-time snowplow phase of a supernova cluster.
\citet{Kipper.etal:25} describes and models the density profile and
time dependence of super bubbles based on \citet{Yadav.etal:17} and
MHD simulations from \citet{Kim.Ostriker:17}.  Their model includes
the evolution of the bubble in both space and time.  For easy of
implementation, we implement a simplified version of their
supper-bubble model as a time-static profile with a temporally
evolving amplitude.

We follow \cite{Kipper.etal:25} in defining an inner and outer wall
radius, \(R_1\) and \(R_2\) respectively, an overall mass, \(M_b\),
and a half-life \(\tau\).  The density for \(R<R_1\) is assumed to be
negative and equal to \(-3M_b(t)/4\pi R_1^3\).  The density for
\(R_1\le R \le R_2\) is assumed to be equal to
\(3M_b(t)/4\pi (R_2^3 - R_1^3)\) to conserve overall disk mass.  The
total duration of the bubble is \(2\tau\) with time dependence
\begin{equation}
  M_b(t, t_o) = M_b
  \begin{cases}
    t - t_o & \mbox{if } 0 <  t - t_o < \tau \\
    2\tau - t + t_o & \mbox{if } \tau < t - t_o < 2\tau \\
    0. & \mbox{otherwise}
  \end{cases}
  \label{eq:bubbletime}
\end{equation}

The algorithm for super bubbles in \EXP\ is:
\begin{enumerate}
\item At every time step, a bubble is generated using a Poisson
  distribution with rate \(f_b\).  We use the value of
  \(f_b = 25\,\mbox{kpc}^{-2}\mbox{Gyr}^{-1}\) from
  \citet{Kipper.etal:25} as our fiducial value.
\item When an event occurs, the bubble center is assigned randomly
  with an exponential disk density distribution with scale length
  \(a\) and random azimuth \(\phi\) at \(z=0\).  The bubble is created
  with time \(t_o=t\) and assigned to the active bubble list.
\item The list of active bubbles is pruned at each time step by
  removing bubbles with \(t>t_o+2\tau\).
\item At each step, the code finds all particles inside any active
  bubble and computes the force for each bubble on the particle based
  on the enclosed mass. The force from the bubble is outward from its
  center.
\item A fixed bubble profile rather than the expanding and contracting
  model \citep{Kipper.etal:25} will over estimate the effect of the
  bubbles but compensates for limitations in spatial resolution.  As a
  test, we also explored evaluating the force by projecting the bubble
  distribution onto the disk expansion basis and allowing the bubbles
  to change their size in time as proscribed in
  \citet{Kipper.etal:25}.  The results from this approach are
  consistent with the direct force evaluation for the time-static
  bubble profile.
\end{enumerate}
We will see in Section \ref{sec:SFmain} that the bubble model for
fiducial parameters negligibly affects bar growth compared to the AGN
feedback model.  For this reason, we abandon further exploration of
the details of this bubble model.

\subsection{A perturbative simulation}
\label{sec:pertsim}

Section \ref{sec:results} shows that application of the AGN
feedback model (Sec. \ref{sec:central}) to the full N-body simulation
described above leads to a suppression of the bar instability for some
ranges of \(\epsilon\) and \(\tau_1\).  To provide explicit dynamical
insight, we present an analysis based on the underlying mechanism of
the bar instability itself.  In particular, \citet{Lynden-Bell:1979}
elegantly derives a condition for an \(m=2\) disturbance to grow in a
galaxy disk.  This idea underpins our understanding of what makes a
bar.\footnote{See the recent paper by
  \citet{Polyachenko.Shukhman:2020} for a description of the mechanism
  and its history} The gist of the idea begins with a decomposition of
a periodic disturbance into its natural harmonic components in
action-angle variables.  The existence of this Fourier-type series
relies on the pure quasi-periodic nature of orbits in the axisymmetric
disk.  Each term in the expansion describes the contribution from a
particular commensurability:
\(l_r\Omega_r + l_\phi\Omega_\Phi = 2\Omega_p\) where \(\Omega_r\) and
\(\Omega_\phi\) are the radial and azimuthal orbital frequencies,
respectively, and \(\Omega_p\) is the bar pattern speed.  By
restricting one's attention to the inner Lindblad resonance (ILR)
which has \(l_r=-1, l_\phi=2\), we get an equation of motion in one
degree of freedom.  A Taylor series expansion of this equation of
motion about the resonant orbit yields Lynden-Bell's celebrated
result: a nascent bar will grow if
\begin{equation}
  {\cal L}\equiv\frac{\partial^2 H_0}{\partial I_s^2} > 0
  \label{eq:LB}
\end{equation}
where \(H_0\) is the Hamiltonian for the background profile and
\(I_s\) is the resonant or \emph{slow} action.  For the ILR, \(I_s\)
is twice the angular momentum, \(L_z\). Equation (\ref{eq:LB}) is
easily evaluated numerically for any regular system. The marginal
growth case for nearly circular orbits occurs for nearly flat but
slightly falling rotation curves. Let the maximum angular momentum at
fixed guiding-center energy be \(J(E)\).  The general trends are that
\({\cal L}\) increases at fixed energy as the relative angular
momentum, \(L_z/J(E)\) decreases, and \({\cal L}\) decreases for fixed
\(L_z/J(E)\) as the guiding center energy (or radius) increases.

The stochastic processes from Section \ref{sec:pert} makes
analytic predictions challenging.  Rather than solve for the system of
stochastic ordinary differential equations directly, we use the hybrid
numerical perturbation theory framework described in a previous paper
\citep{Weinberg.Katz:2007} to obtain a particle system.  In essence,
the radial motion is much faster than the libration period of the
orbit near resonance.  The fast motion is therefore adiabatically
invariant to changes on the scale of the libration period.  This
allows us to use the averaging principle to isolate the dynamics
controlled by the ILR.  This leaves a one-dimensional Hamiltonian
described the precession angle, \(w_s\), and slow action, \(I_s\), for
some effective quadrupole potential describing the full perturbation.

The overall perturbation felt by one phase-averaged particle depends
on sum of all other phase-averaged contributions to the effective
potential.  The density of these phase-averaged particles are
quadrupoles that look like dumbbells.  The Hamiltonian perturbation
theory describes precisely how an ensemble of dumbbells interact.  We
employ the same mean-field ideas that underlie the BFE N-body methods
from \EXP\ (see Section \ref{sec:exp}).  Specifically, each dumbbell
particle makes a contribution to the mean field of the time-averaged
quadrupole potential that is represented by a vector of basis-function
coefficients.  Appendix \ref{sec:Hpert} derives this formalism and
describes the numerical implementation of the method which we call
\emph{BarHMF}.

In the linear limit, the method is a subset of standard matrix
response theory used to identify instabilities.  It is a subset in the
sense that it only includes one or several commensurabilities
\((l_r, l_\phi)\) while the general theory includes many.  Conversely,
this hybrid N-body scheme is itself non-linear: the coupling between
phase-averaged particles take the form of coupled non-linear pendula.
Therefore, the interaction kernel between all pairs of particles is
not limited to linear excitations.  For example, the BarHMF simulation
demonstrates the expected exponential growth and non-linear saturation
of a traditional N-body bar.  The dynamics includes trapping effects
that are not part of the standard linear response theory.  In this
sense, we have constructed a particular non-linear analog model that
matches linear perturbation theory where it is valid and exhibits many
of the features of non-linear bar growth.  We will apply the
stochastic feedback from Section \ref{sec:exp} to this idealized bar
model to help explain the observed dynamics of the full N-body
simulations with AGN feedback.

\begin{figure*}
  \mbox{
    \includegraphics[width=0.49\textwidth]{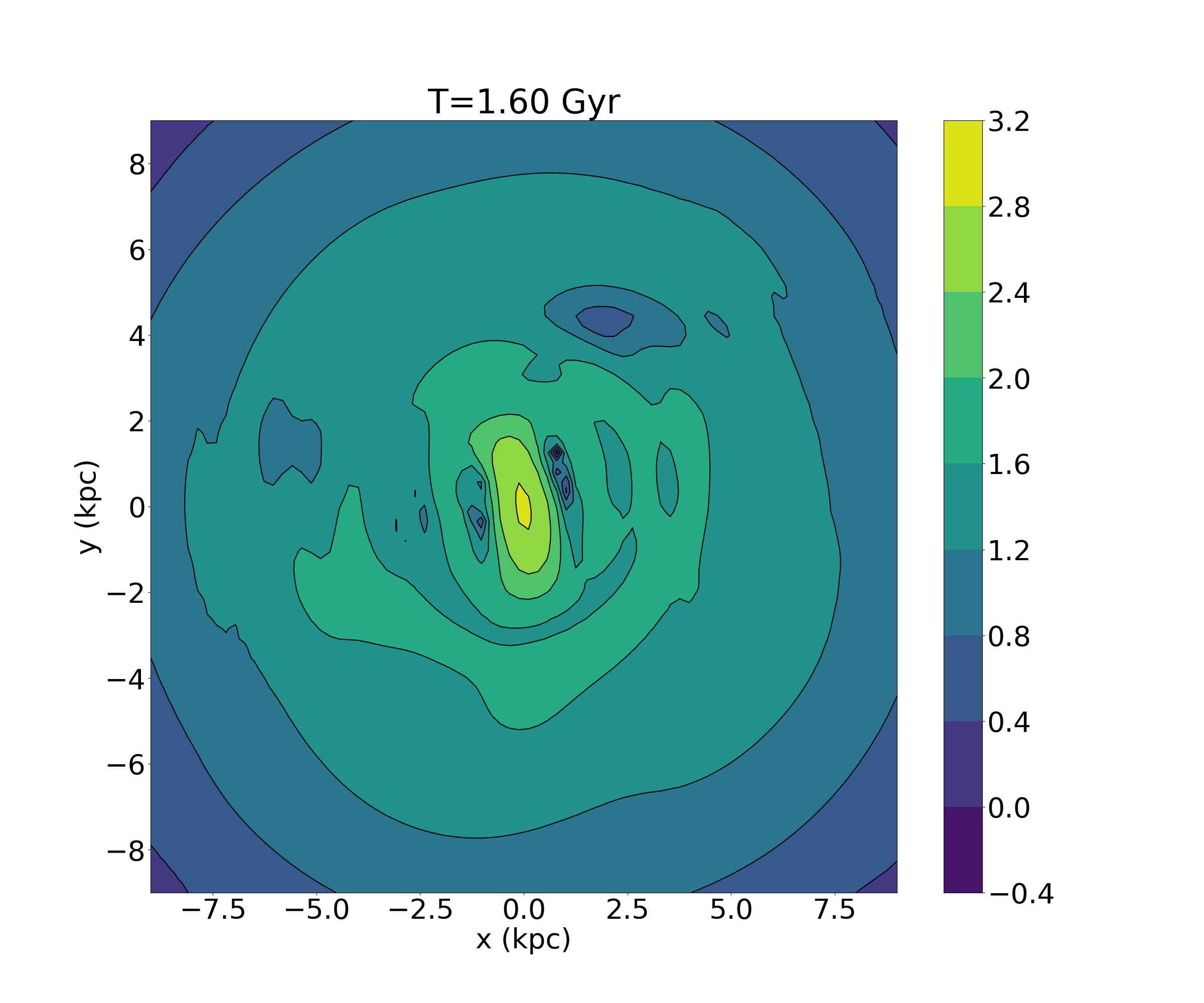}
    \includegraphics[width=0.49\textwidth]{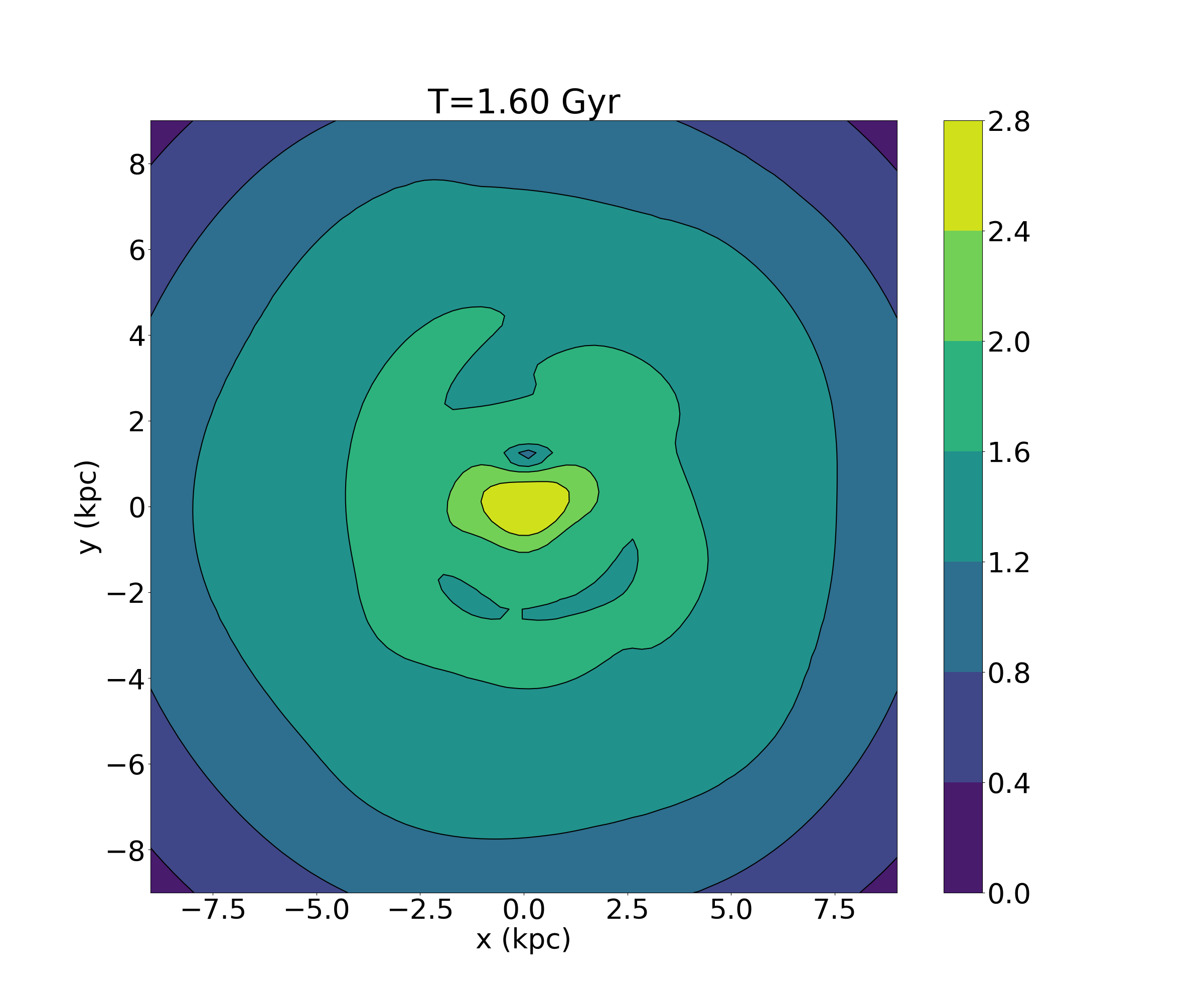}
  }
  \mbox{
    \includegraphics[width=0.49\textwidth]{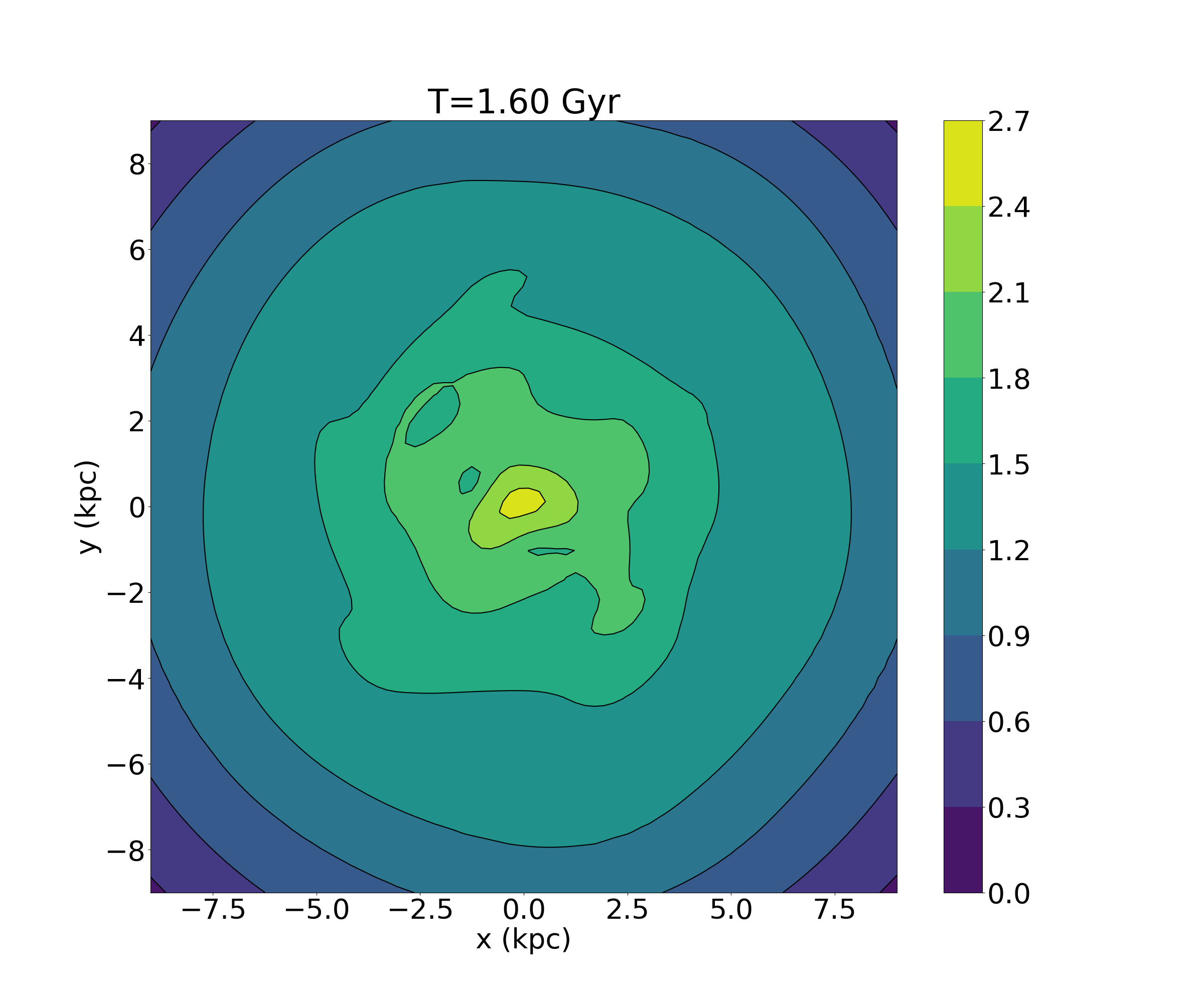}
    \includegraphics[width=0.49\textwidth]{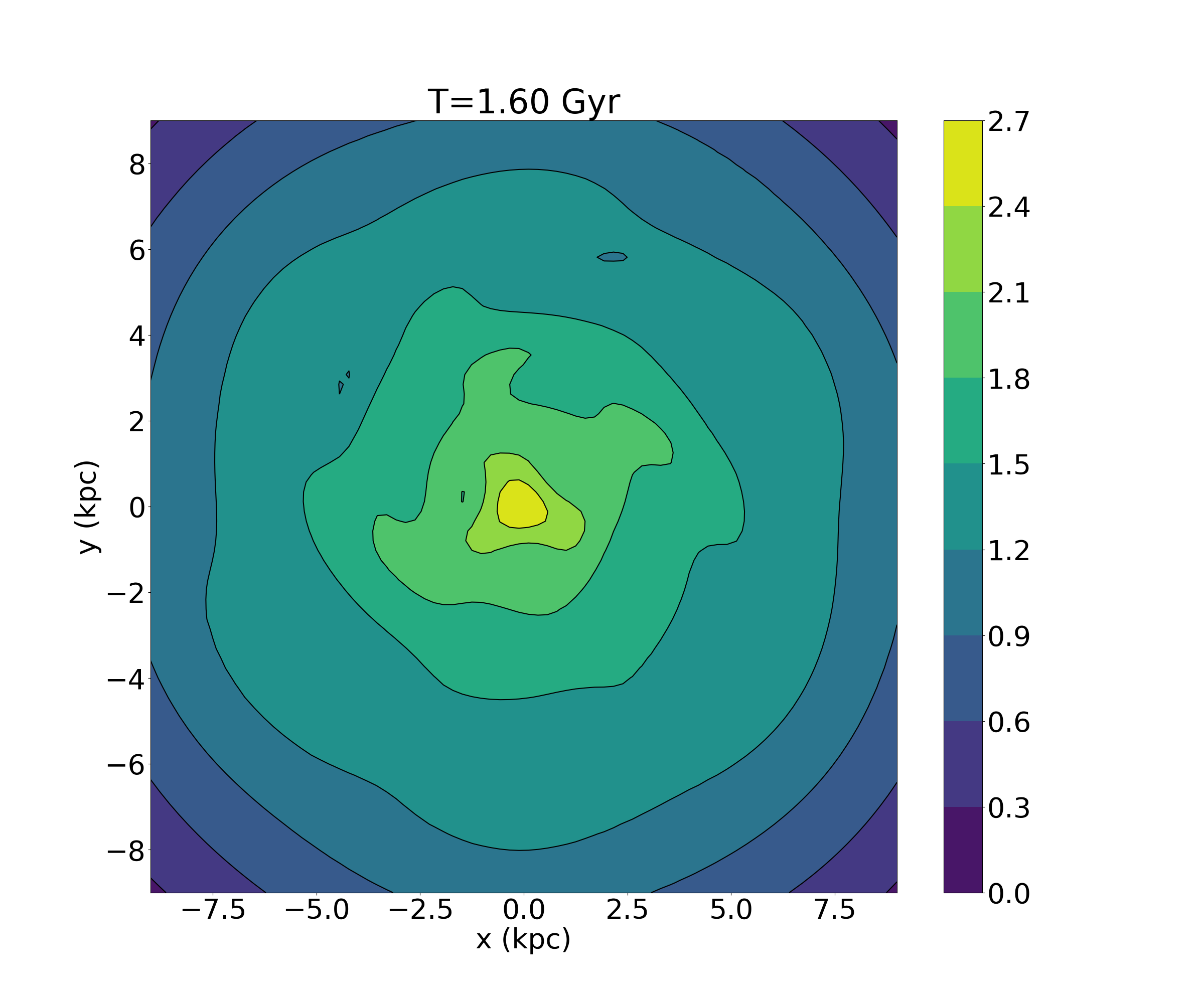}
  }
  \caption{\label{fig:densDemo}The disk surface density at
    \(T=1.6\Gyr\).  The upper-left panel shows the primary model with no
    central feedback.  The next three panels have central feedback
    with \(\epsilon=0.1\) and \(\tau_1=20, 40, 80\) Myr and
    \(\tau_2=\tau_1/2\), from left-to-right and top-to-bottom.  The
    density scale is logarithmic in units of \(\msun/\pc^2\).}
\end{figure*}

\begin{figure*}
  \mbox{
    \includegraphics[width=0.49\textwidth]{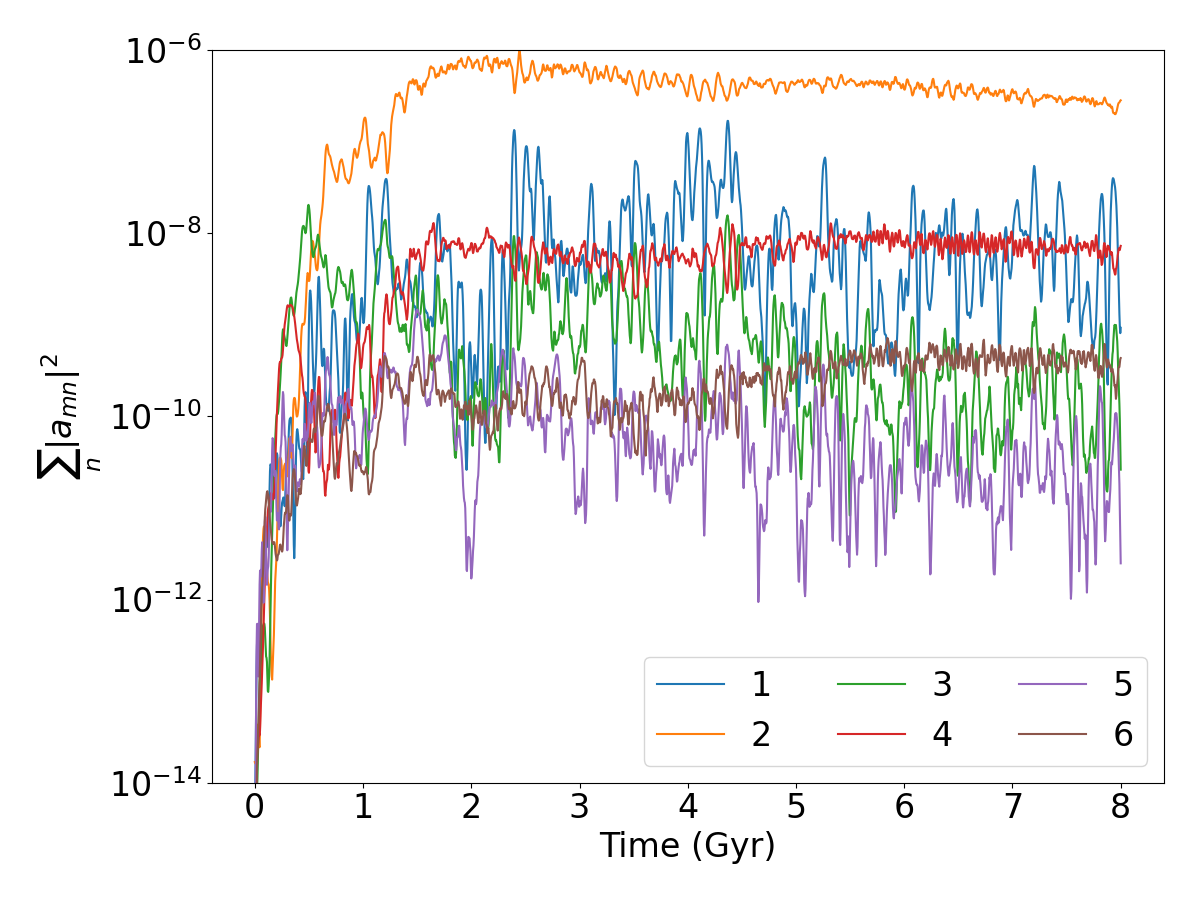}
    \includegraphics[width=0.49\textwidth]{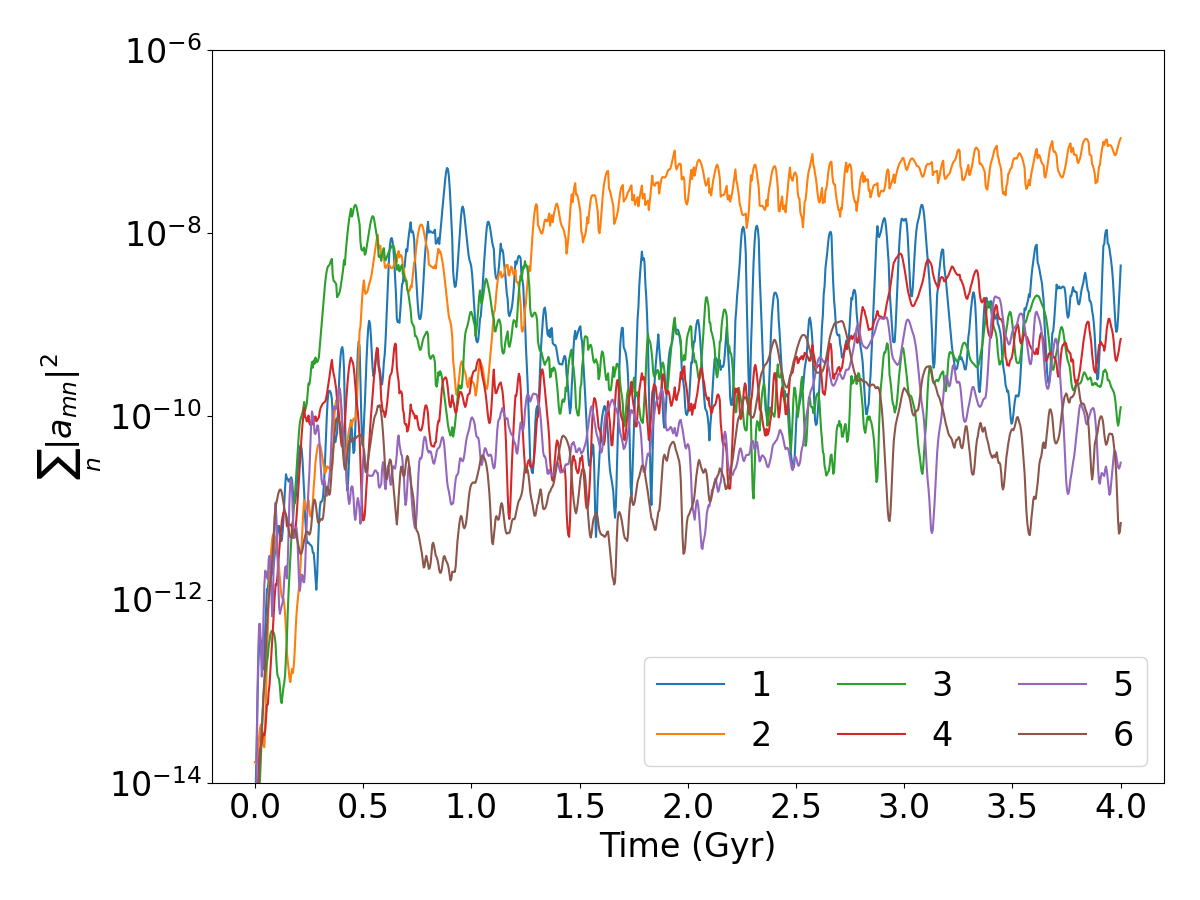}
  }
  \mbox{
    \includegraphics[width=0.49\textwidth]{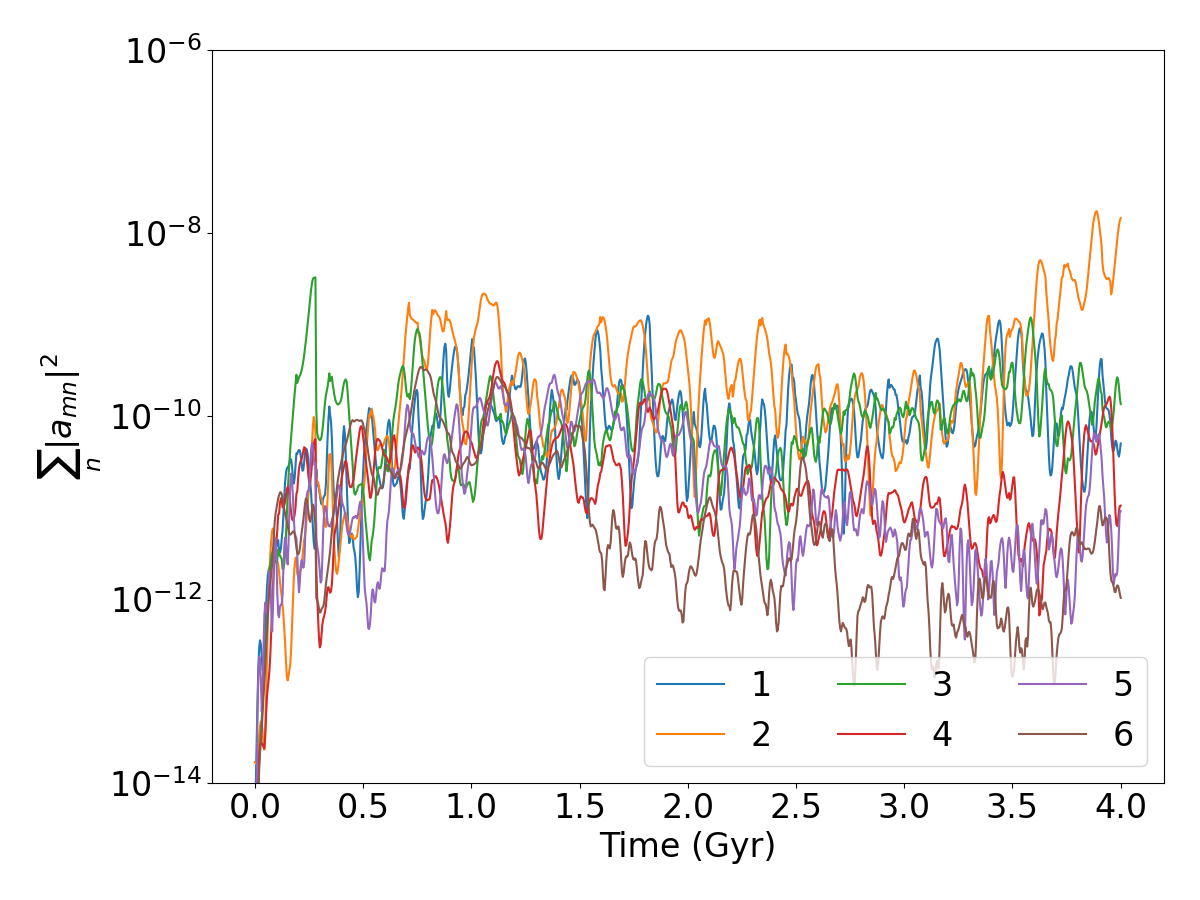}
    \includegraphics[width=0.49\textwidth]{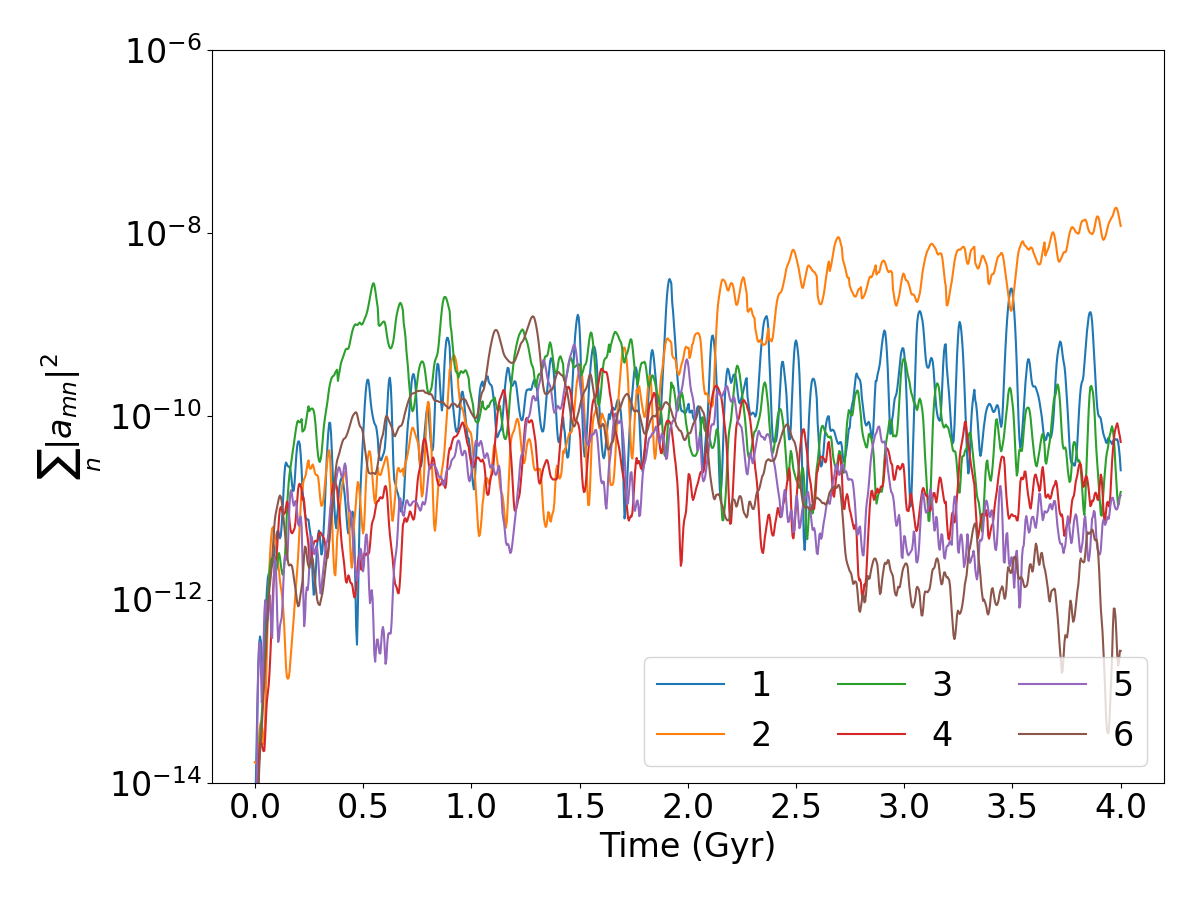}
  }
  \caption{\label{fig:powerDemo}The gravitational power for azimuthal
    harmonics \(m=1,\ldots,6\) (see inset legends) for each of the
    four runs from Fig. \ref{fig:densDemo}. The power had units of
    virial energy, \(G M_{vir}^2/R_{vir}\).  The \(m=5,6\) harmonics
    have very little correlated power and are a good indicator of the
    particle noise floor.}
\end{figure*}

\section{Results}
\label{sec:results}

We begin in Section \ref{sec:AGNmain} with a summary demonstration of
the consequences to bar formation with a series of 4 runs with
successively more AGN central feedback as described in Section
\ref{sec:central}.  We compare this to bar formation with no feedback
in Section \ref{sec:primary}.  Section \ref{sec:SFmain} shows that bar
formation is not sensitive super bubble formation using the model from
Section \ref{sec:super} unless the outflows are unrealistically large
and centrally concentrated, much like the AGN model.  Later sections
describe the sensitivity of the bar interaction to the time scale and
amplitude of the AGN feedback process.  We end, in Section
\ref{sec:barhmf} with dynamical insight from the BarHMF model.

\subsection{AGN feedback: a quick demonstration}
\label{sec:AGNmain}

A typical central-engine event may be short lived,
\(\lesssim10^{4}\mbox{--}10^{5}\,\mbox{yr}\)
\citep{Schawinski.etal:10}.  This is \emph{sudden} from the orbital
viewpoint.  However, simulations suggest and observations
indicate mass flow events on longer time scales, \(\sim1\Myr\)
\citep{Zubovas.etal:23}. Observed bubbles in radio galaxies suggest
blowout events with cadence of tens of millions of years
\citep{Biava:21}.  Even longer upper limits follow from considering
the duty cycle of observed AGN activity in the Universe assuming that
all galaxies have SMBHs. These duty cycles are likely to have been
larger at cosmic noon.  In summary, estimates of burst frequency vary
from \(10\Myr\) to several \Gyr. In our stochastic model, \(\tau_1\)
is the average time between central events, not the event duration.
The key feature is that these burst time scales are comparable to the
characteristic dynamical time scales.

For this initial demonstration, we consider four runs of the primary
model: the first run has no feedback, the second three fix
\(\epsilon=0.1\) with \(\tau_1=20, 40, 80\Myr\) from the feedback
prescription described in Section \ref{sec:pert}. We explore a larger
range for \(\tau_1\) in Section \ref{sec:tau}. The results are
compared in Figure \ref{fig:densDemo}.  We adopt \(\tau_2 = \tau_1/2\)
in equations (\ref{eq:Pevent}) and (\ref{eq:Mdecay}) which allows the
central disk to recover its lost mass on a time scale shorter than the
average interval between outflow events.  We have tried other choices,
such as \(\tau_2=\tau_1/4\) and \(\tau_2=\tau_1\) and the results are
qualitatively unchanged.  The key dynamical effect is the
decorrelation of the orbital precession that is induced by feedback
prescription.

The strength of the non-axisymmetric features are nicely described by
total power in each azimuthal harmonic \(m\).  The biorthogonal
functions used in the BFE potential solver naturally describe this
power. Specifically, the sum of the moduli for all coefficients with a
particular value of \(m\) is twice the negative of the total energy in
the gravitational field at harmonic order \(m\); we define this
positive quantity as gravitational power.  The square root of this
power measures the mean amplitude of the gravitational potential in
the BFE \citep{Petersen.etal:2022}.  Figure \ref{fig:powerDemo} show
the traces of the power for each of the non-axisymmetric orders
\(m\in[1,6]\) for the runs described in Figure \ref{fig:densDemo}.  In
the case without feedback (upper-left panel of
Fig. \ref{fig:powerDemo}), we clearly see coherent signal at all of
the even harmonics \(m\) with the amplitude of \(m=4\) about 0.1 of
\(m=2\).  The \(m=4\) is responsible for the bar's \emph{boxiness}.
Both dynamics of the bar and the natural modes in the dark-matter halo
cause \(m=1\) distortions.  These are tend to fluctuate but have
similar amplitude to the \(m=4\) signal.  Upon including central
feedback, the bar growth is strongly suppressed for all three values
of \(\tau_1\).  The \(\tau_1=20, 80 \Myr\) cases reveal some hint of
an \(m=2\) bar-like feature but comparison with the no feedback run
reveals that these are down by an order of magnitude from the no
feedback run (upper-left panel).  The \(\tau_1=40\Myr\) case has no
obvious bar.  The minimum at \(\tau_1=40\Myr\) is naturally explained
by increased coupling between the Markov jump process and the orbital
frequencies necessary to support the bar.  We will demonstrate this
further in Section \ref{sec:barhmf}.

\subsection{Evolution without feedback}
\label{sec:primary}

\begin{figure*}
  \mbox{
    \includegraphics[width=0.49\textwidth]{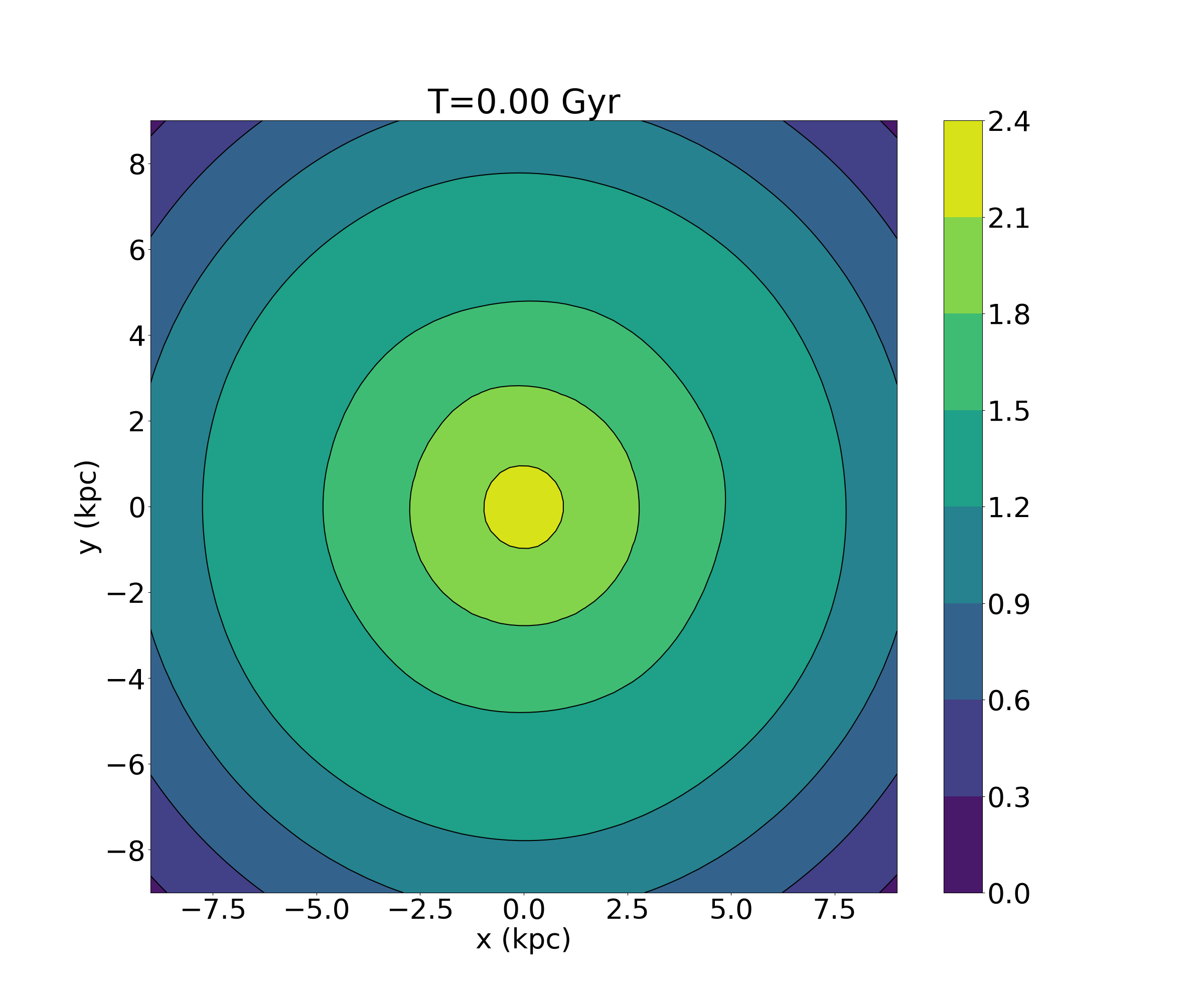}
    \includegraphics[width=0.49\textwidth]{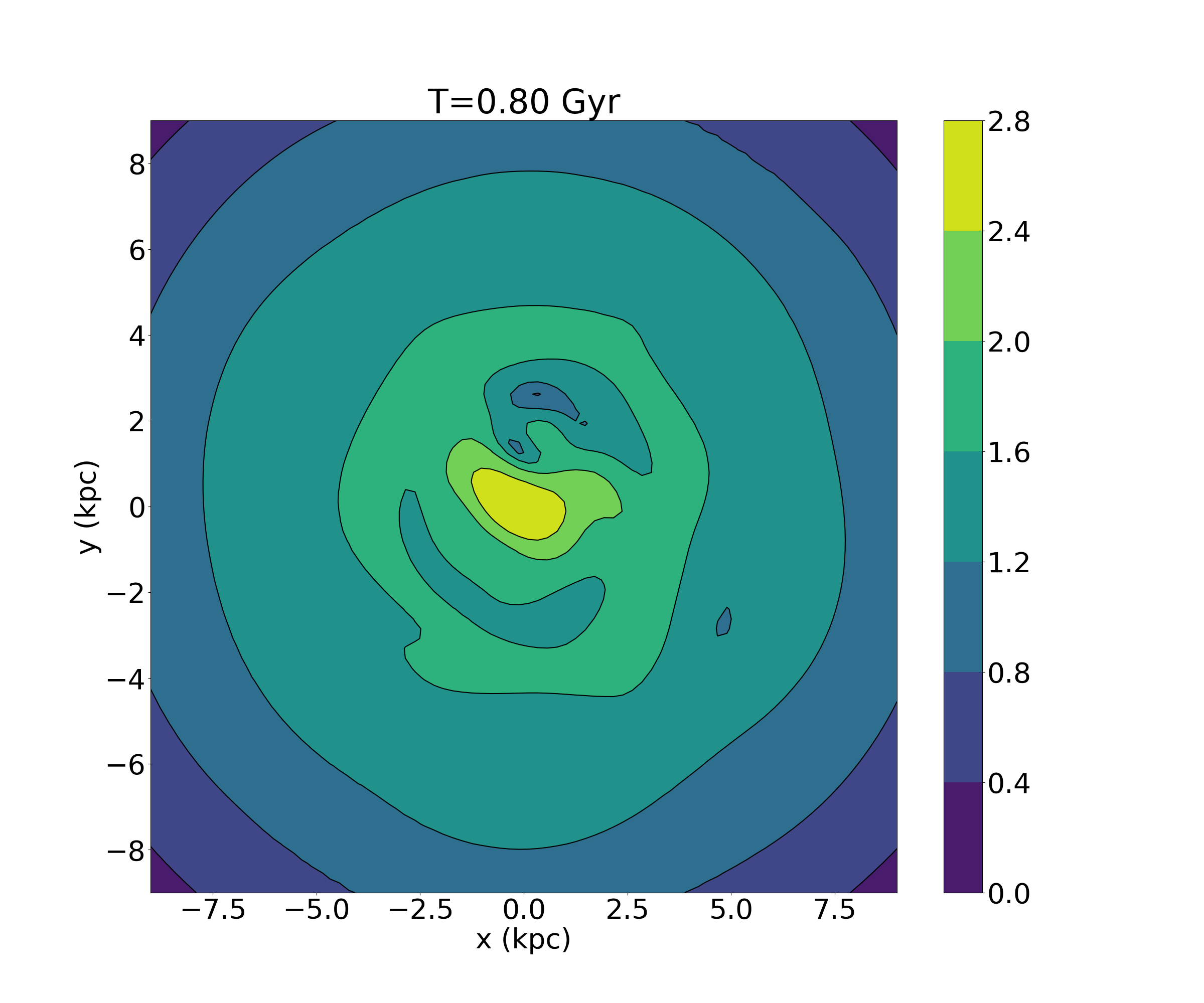}
  }
  \mbox{
    \includegraphics[width=0.49\textwidth]{run125_4r_1p60.png}
    \includegraphics[width=0.49\textwidth]{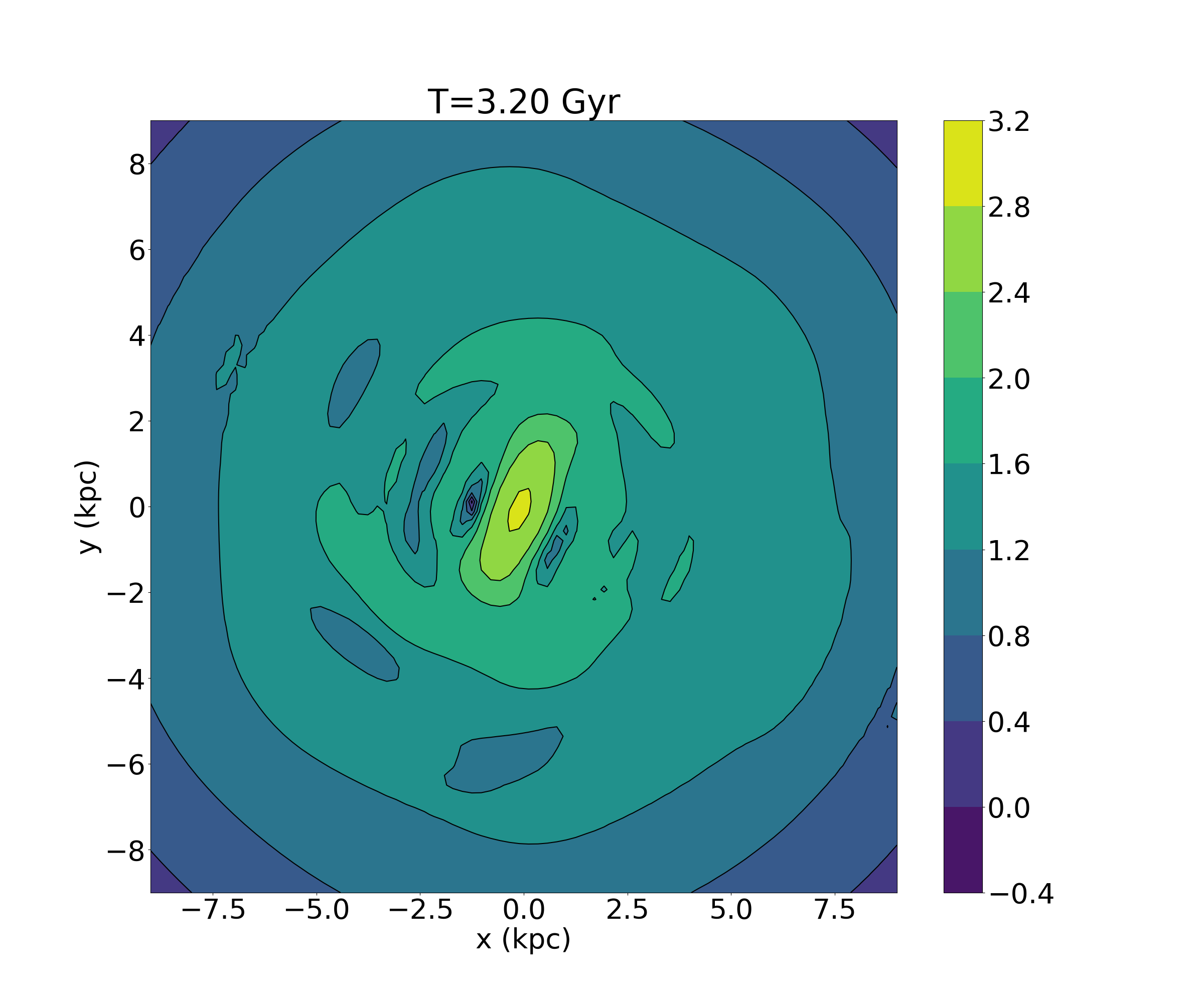}
  }
  \caption{\label{fig:densfid}The surface density for the
    \emph{primary} model with no feedback (\(\epsilon=0\)) at T=0,
    0.8, 11.6, 3.2 Gyr.  The bar reaches steady state by \(T=2\Gyr\).  The
    density scale is logarithmic in units of \(\msun/\pc^2\).}
\end{figure*}

For comparison, we present a brief description of full N-body
evolution using the primary model from Section \ref{sec:model}
\emph{without} any feedback.  This provides a reference for comparing
the bar growth and evolution to simulations \emph{with} feedback.  The
salient features of the evolution without feedback are as follows:
\begin{enumerate}
\item The growing bar becomes distinct at \(T\approx600\) Myr.
\item The pattern speed slows quickly during formation as it grows in
  strength and lengthens.
\item The bar reaches an approximate steady-state at \(T\approx 2\)
  Gyr.  The bar pattern speed decays slowly and nearly linearly
  thereafter.
\end{enumerate}

\begin{figure}
  \includegraphics[width=0.5\textwidth]{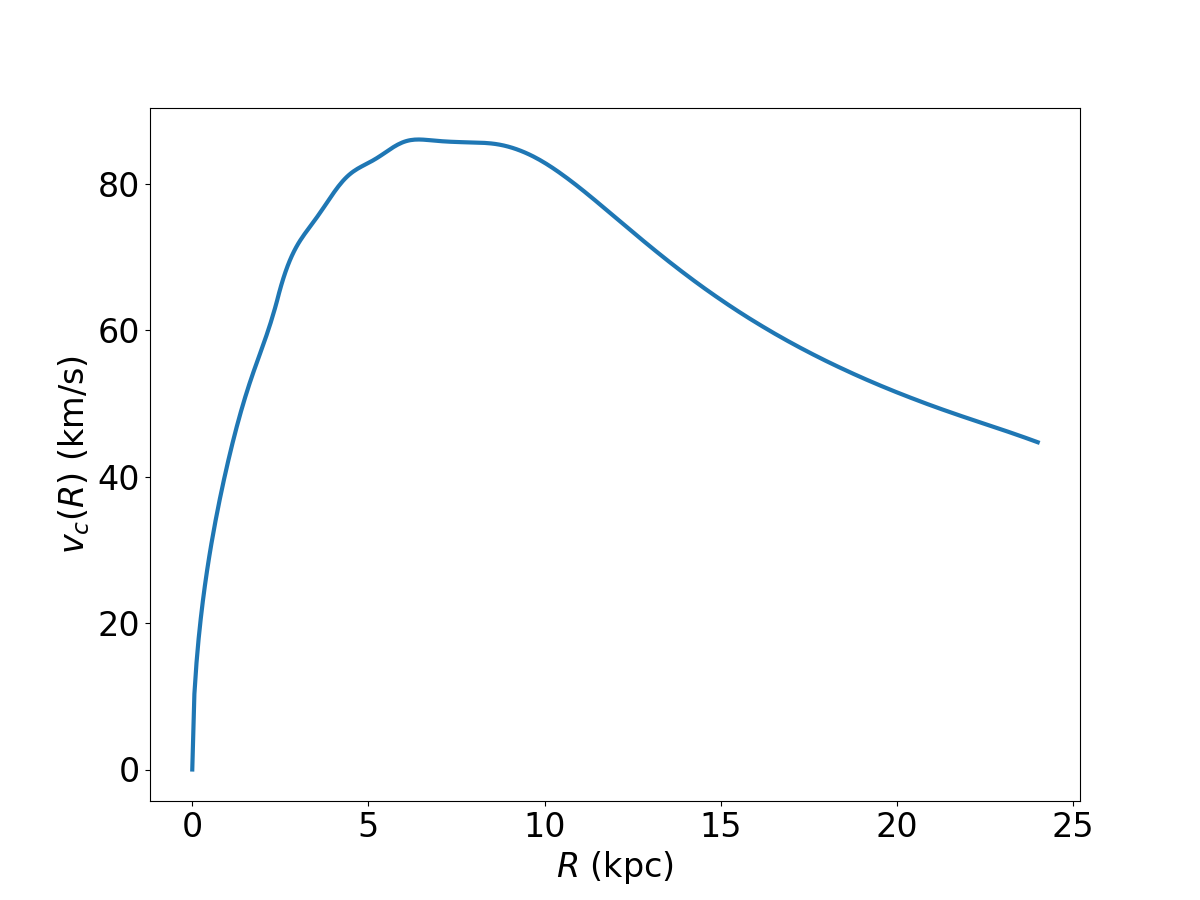}
  \caption{\label{fig:rotcurve} The circular velocity curve for the
    unperturbed halo and disk model described in Section
    \ref{sec:disk} computed from the radial force in the simulation at
    \(T=0\).  The curve is approximately linearly increasing in the
    first scale length, \(R=3\kpc\), with a peak at \(R\approx5\kpc\)
    and a slow decline thereafter.}
\end{figure}

Figure \ref{fig:densfid} shows the surface density at the start and at
\(T=1, 2, 4\) Gyr.  The first panel shows the initial conditions
reconstructed from the BFE.  The second panel shows the bar at the of
its growth phase.  The bar is very slowly evolving in third and fourth
panels.  The rotation curve for the initial model
(Fig. \ref{fig:rotcurve}) shows the classic linearly rising inner
profile that promotes bar growth following the Lynden-Bell condition
(eq. \ref{eq:LB}).

\begin{figure}
  \includegraphics[width=0.45\textwidth]{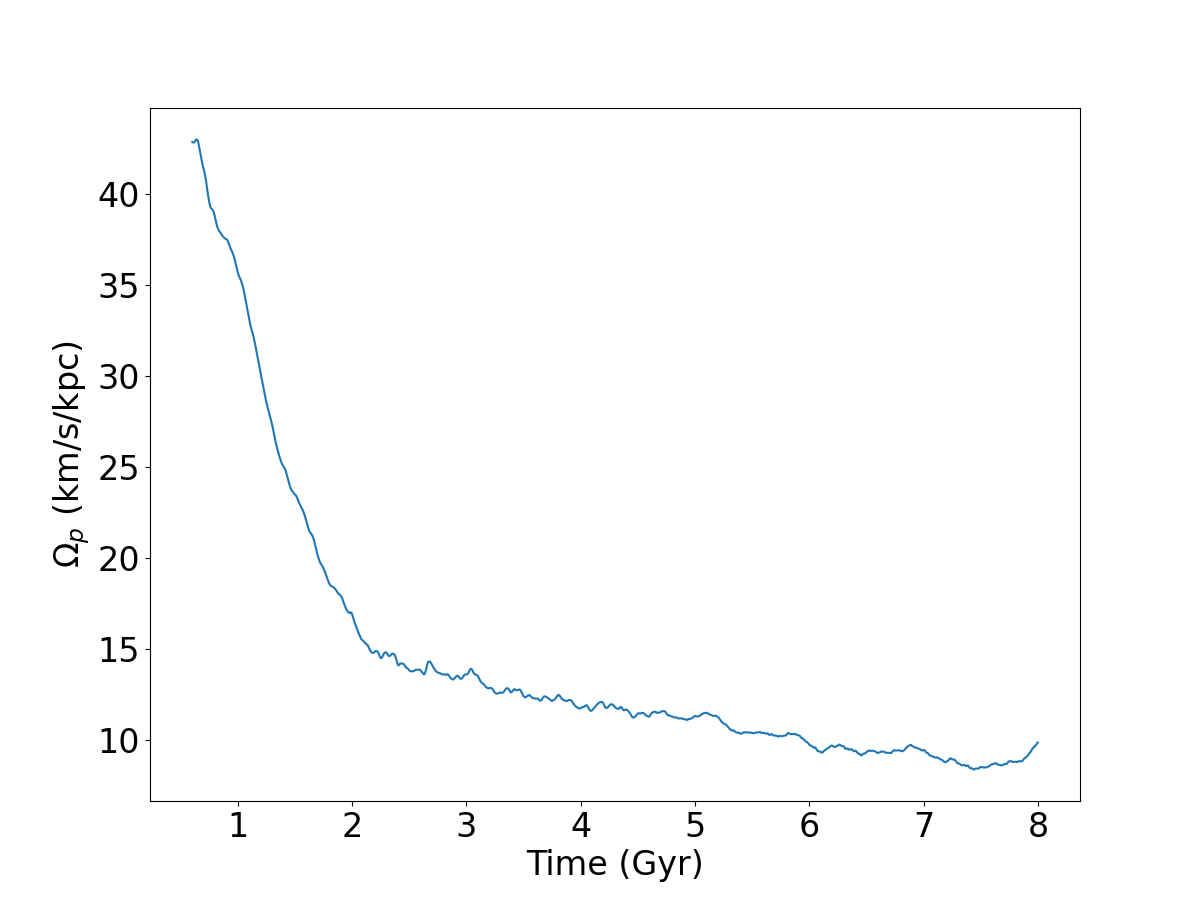}
  \caption{\label{fig:pattern}Bar pattern speed for the primary run
    shown from the point of bar assembly onward. }
\end{figure}

Figure \ref{fig:pattern} describes the pattern speed, \(\Omega_p\),
which is computed from the complex phase of the first three radial
terms in the \(m=2\) part of the BFE.  These first three terms contain
80\% of the total \(m=2\) power and accurately portray the shape and
potential of the bar.  The rapid change in \(\Omega_p\) for
\(600\Myr\lesssim T\lesssim 2\Gyr\) corresponds to the rapid growth in
\(m=2\) power.  In most of this discussion, we will only consider
evolution up to \(T=4\) Gyr given our \emph{cosmic noon} perspective;
we show the power in the first panel of Figure \ref{fig:powerDemo} and
pattern speed in Figure \ref{fig:pattern} up to \(T=8\) Gyr for
completeness.  This run of pattern speed with time compares well with
the primary run from \citet{Petersen.etal:2019} providing a check for
the two-dimensional basis functions described in Section \ref{sec:exp}
and Appendix \ref{sec:2ddisk}.  In summary, the orbits driving the bar
sheds angular momentum as the bar amplitude grows.  These orbits
become self-trapped in the quadrupole potential, increasing their
eccentricity, lengthening the bar, and decreasing the pattern speed.
After the exponential growth phase ceases, the bar continues to grow
in mass and length by trapping orbits at the end of the bar and
shedding angular momentum to the disk and DM
halo. \citet{Petersen.etal:2019} called this the \emph{secular growth
  phase}.  Without a source new stars or gas, the pattern speed
continues to slowly drop and the bar continues to slow grow in mass
and length.

\subsection{Results for SF feedback}
\label{sec:SFmain}

Our simulations suggests that AGN feedback creates a global stochastic
impulse in the gravitational field that couples to the orbital
dynamics of the forming bar.  This is corroborated with a dynamical
mechanism in Section \ref{sec:barhmf}.  However, star-formation
outflow is spatially distributed in the disk and each outflow event is
individually less massive rather than a centrally concentrated AGN
event.  In this section, we use the super-bubble model from Section
\ref{sec:super} to explore whether SF outflow has a measurable effect
on bar growth.  This model assumes that super bubbles formed from
coalescing individual SN outflows drive outflow events at multiple
sites across the disk.

Our fiducial bubble model has \(R_i=690\pc\), \(R_o=750\pc\),
\(M_b=1.2\times10^{4}\msun\) and half-life \(T=40\Myr\).  The bubble
surface density estimate from \citet{Kipper.etal:25} for a saturated
star burst is \(f_b\approx 25\,\mbox{kpc}^2\mbox{Gyr}^{-1}\).  We
translate this into a rate of bubble production inside of scale length
of the disk as described in Section \ref{sec:super}.

A simulation with fiducial parameters has negligible effect on the
forming bar.  We followed this fiducial simulation with a suite of 54
additional simulations with varying combinations of key parameters in
turn.  We find that the following changes have negligible effect: (i)
doubling the bubble radii \(R_i\) and \(R_o\) for fixed bubble mass
\(M_b\), half-life and rate; (ii) increasing the half-life by a
factors of 2, 4, and 8 (iii) increasing the bubble mass \(M_b\) by a
factor of 10, 100, and 1000.  Conversely, we found that concentrating
the bubbles in the center of the galaxy and increasing the mass of the
bubbles by factors of 100 or 1000 can slow or eliminate the bar
formation. In other words, when the parameters of the SF model become
similar to the AGN model, the same results obtain.  In summary,
astronomically realistic spatially-distributed SF feedback washes out
the correlated signal in the gravitational field fluctuations.  While
the process provides a background noise, it appears unlikely to
generate large enough gravitational potential fluctuations to couple
to the dynamics of bar formation.

The mechanism proposed in this paper couples the bar dynamics to
large-scale fluctuations in the gravitational potential.  However, our
bubbles only reach several scale heights above the disk for
\(R_o = 750\pc\), so perhaps the lack of a significant effect on bar
evolution is not surprising.  We have not investigated the work done
by galactic fountain generation, and this will be the subject of a
future investigation.  Meanwhile, the simple AGN feedback mechanism
from Section \ref{sec:AGNmain} also applies to a strong galactic
fountain in the inner galaxy.  It is the potential fluctuation induced
by the mass outflow over a kiloparsec scale, not the loss of the mass
as a wind, that causes bar suppression or disruption.

\subsection{Sensitivity to the details of AGN feedback}

Having eliminated the SF-driven super bubble feedback model as a
plausible influence on bar formation, we return to the AGN feedback
model.  We quantify the variation of the response with the parameters
of the AGN feedback process by considering two distinct scenarios:
\begin{enumerate}
\item Fixed \(\tau_1=20\Myr\) and varying amplitude of the mass loss,
  \(\epsilon\) in Section \ref{sec:eps}.
\item Fixed \(\epsilon=0.3\) and varying the value of \(\tau_1\) in
  Section \ref{sec:tau}.
\end{enumerate}
While the duration of the AGN energy injection may be shorter than
\(\tau_1\), and a wind crossing time is \(\sim1\mbox{Myr}\), the
relevant time scale for dynamical coupling is the response of the
system to the sudden change in the gravitational potential induced by
the outflow.  This occurs on the dynamical time scales for the
surrounding galaxy.  These timescales are tens to hundreds of Myr.

\subsubsection{Comparison of frequencies}
\label{sec:eps}

\begin{figure}
  \includegraphics[width=0.5\textwidth]{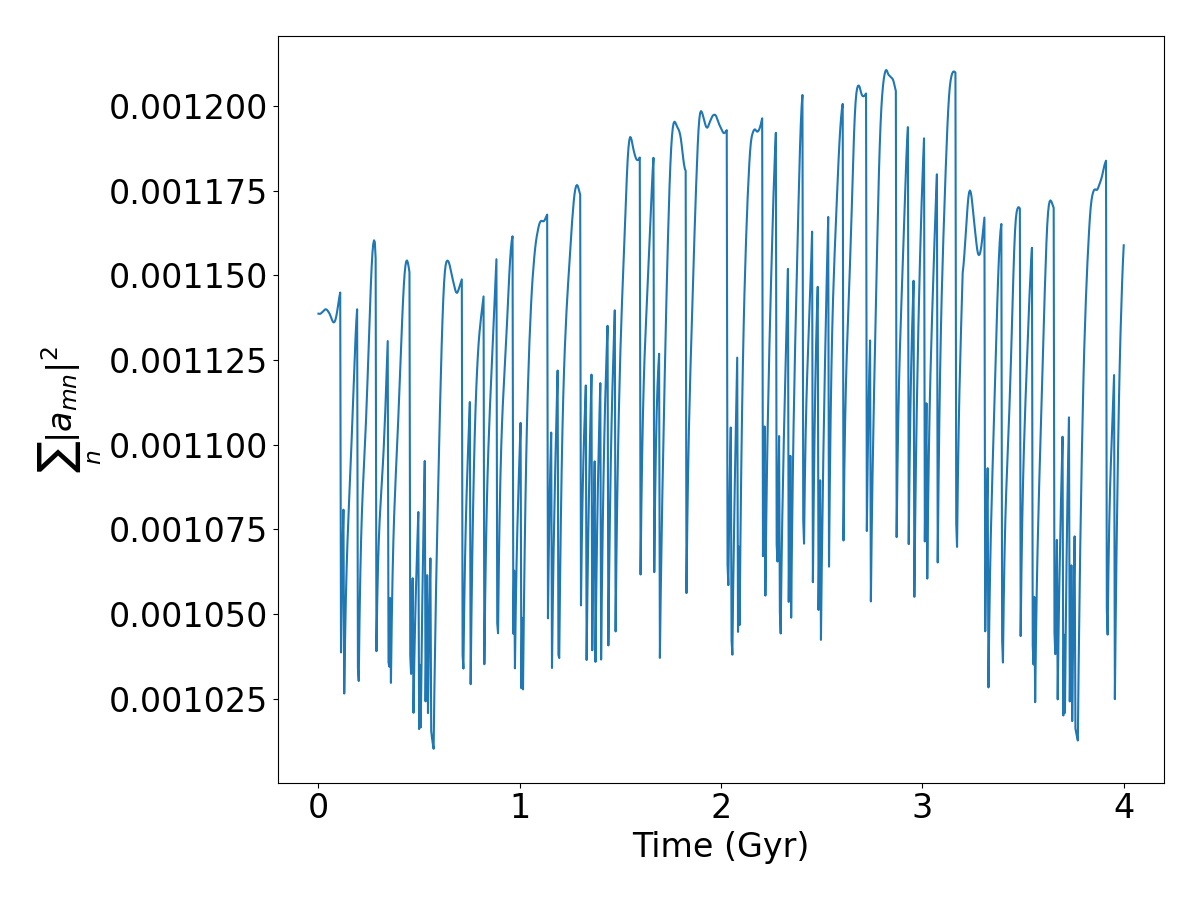}
  \caption{\label{fig:m0}The run of \(m=0\) power for the
    \(\epsilon=0.1\), \(\tau_1=40\) Myr simulation to illustrate the
    fluctuation in the gravitational power induced by star formation
    events.  See Fig. \ref{fig:powerDemo} for an illustration of the
    non-axisymmetric power in the same simulation.}
  \bigskip
\end{figure}

We fix \(\epsilon=0.1\); equation (\ref{eq:deltaM}) implies that the
mass that temporarily leaves the disk per outflow event is 1.4\% of
the total disk mass (\(7\times10^7\msun\)).  This is consistent with
observed estimates \citep{Santoro:20} over 10 Myr.

The \(m=0\) component represents most of the gravitational power in
the disk and is not shown in Figure \ref{fig:powerDemo} to emphasize
the non-axisymmetric components.  We illustrate the effect of
star-formation feedback on the gravitational power of the \(m=0\)
component alone in Figure \ref{fig:m0} for the \(\tau_1=40\Myr\) run.
The value \(\epsilon=0.1\) affects the total gravitational potential
energy of the disk at a level of approximately 0.8\%.

\subsubsection{Comparison of amplitudes}
\label{sec:tau}

\begin{figure}
  \centering
  \includegraphics[width=0.45\textwidth]{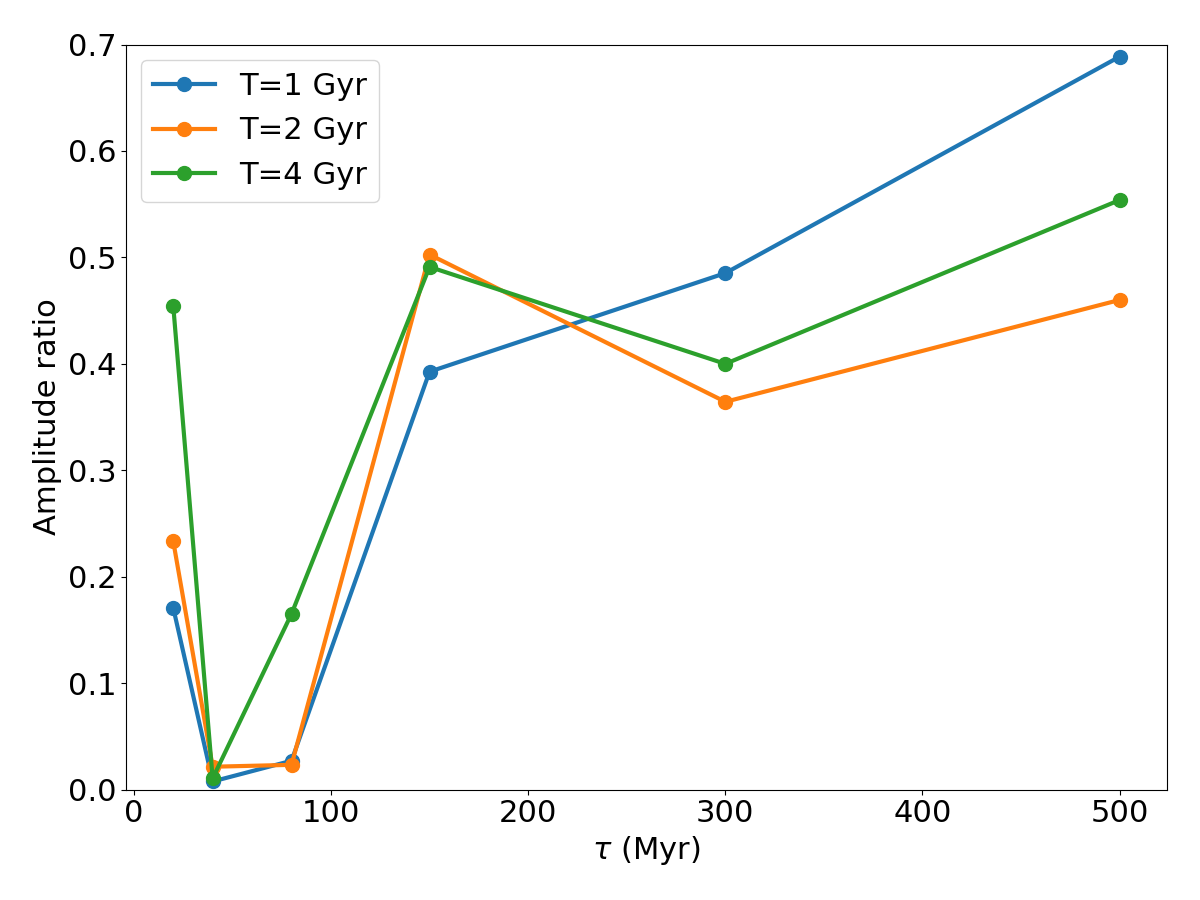}
  \includegraphics[width=0.45\textwidth]{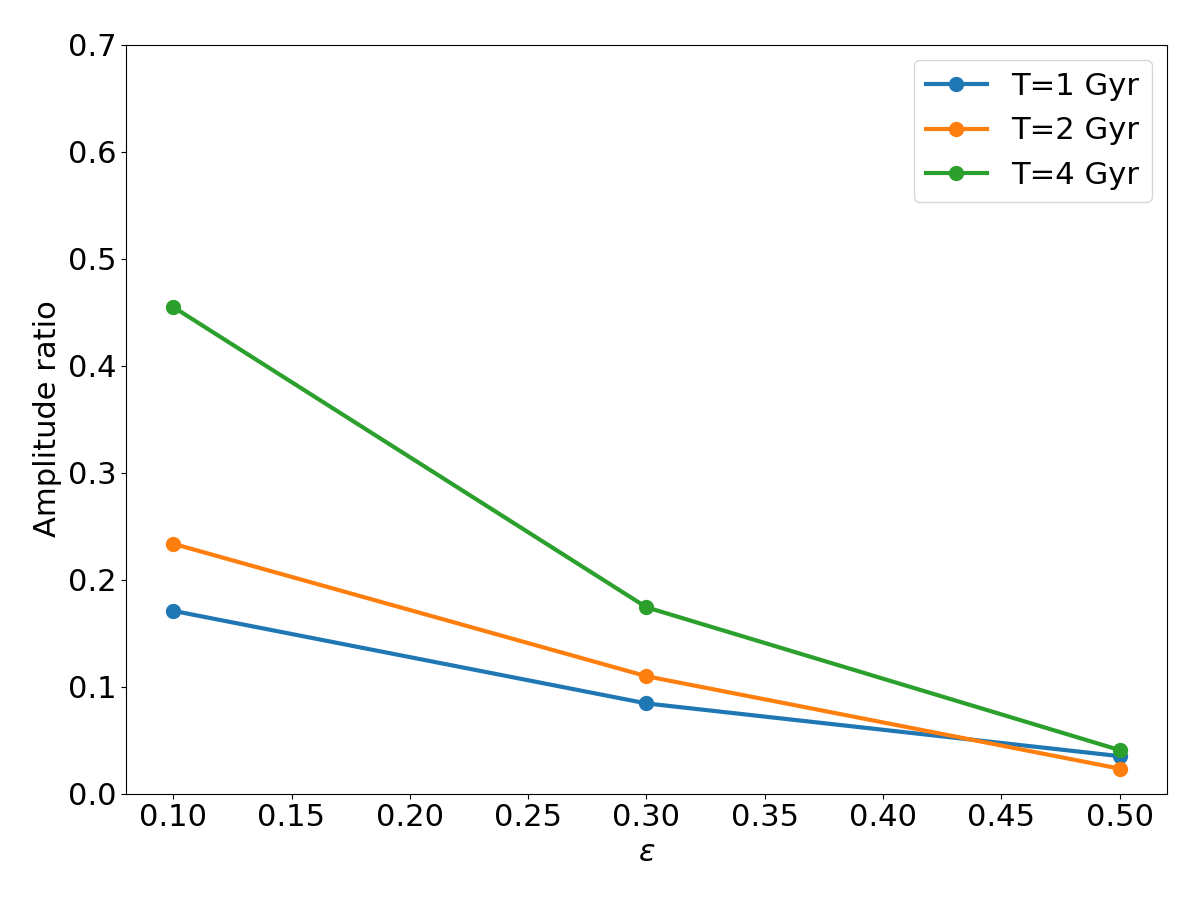}
  \caption{\label{fig:ratios}The ratio of the \(m=2\) amplitudes at
    \(T=1, 2, 4\Gyr\) in simulations with AGN feedback to the
    \(m=2\) amplitude from the primary simulation without feedback.
    Top: constant value of \(\epsilon=0.1\) with values of \(\tau_1\)
    along the x-axis.  Bottom: constant value of \(\tau_1=20\Myr\) with
    values of \(\epsilon\) along the x-axis.}
\end{figure}

Next, we fix \(\tau_1=20\Myr\) and examine the variance with
\(\epsilon\in[0.1, 0.3, 0.5]\). Equation (\ref{eq:deltaM}) implies
that the mass lost per outflow event for these values of \(\epsilon\)
are 1\%, 4\%, and 7\%, respectively.  These results are summarized and
compared with results from previous section in Figure
\ref{fig:ratios}.  This figure shows the ratio of the BFE amplitudes
(root power) at \(m=2\) for the varied values of \(\tau_1\) and
\(\epsilon\) with those from the primary model without feedback at
\(T=1,\,2,\,4\Gyr\).  As expected, the lower panel shows that larger
values of mass loss, \(\epsilon\), lead to smaller amplitude ratios at
all times.  The upper panel summarizes the results from the previous
section with an extended range of \(\tau_1\) from 10--500 Myr: the
largest damage is done to the bar instability for
\(\tau_1\approx50\Myr\).  This time interval is close to the
characteristic orbital time for inner bar orbits.  As \(\tau_1\)
increases, the amplitude ratio increases.  However, even for large
value of \(\tau_1=500\Myr\), the bar amplitude is suppressed by a
factor of two.  To index the amplitude ratio to bar morphology, the
second panel from Figure \ref{fig:densDemo} is an amplitude ratio of
0.23.  The next two panels are ratios of 0.13 and 0.03.  The 0.23 case
(second panel) is typical of a weak bar and ratios below \(\sim0.2\)
(third and fourth panels) are not bar-like.

\begin{figure}
  \centering
  \includegraphics[width=0.45\textwidth]{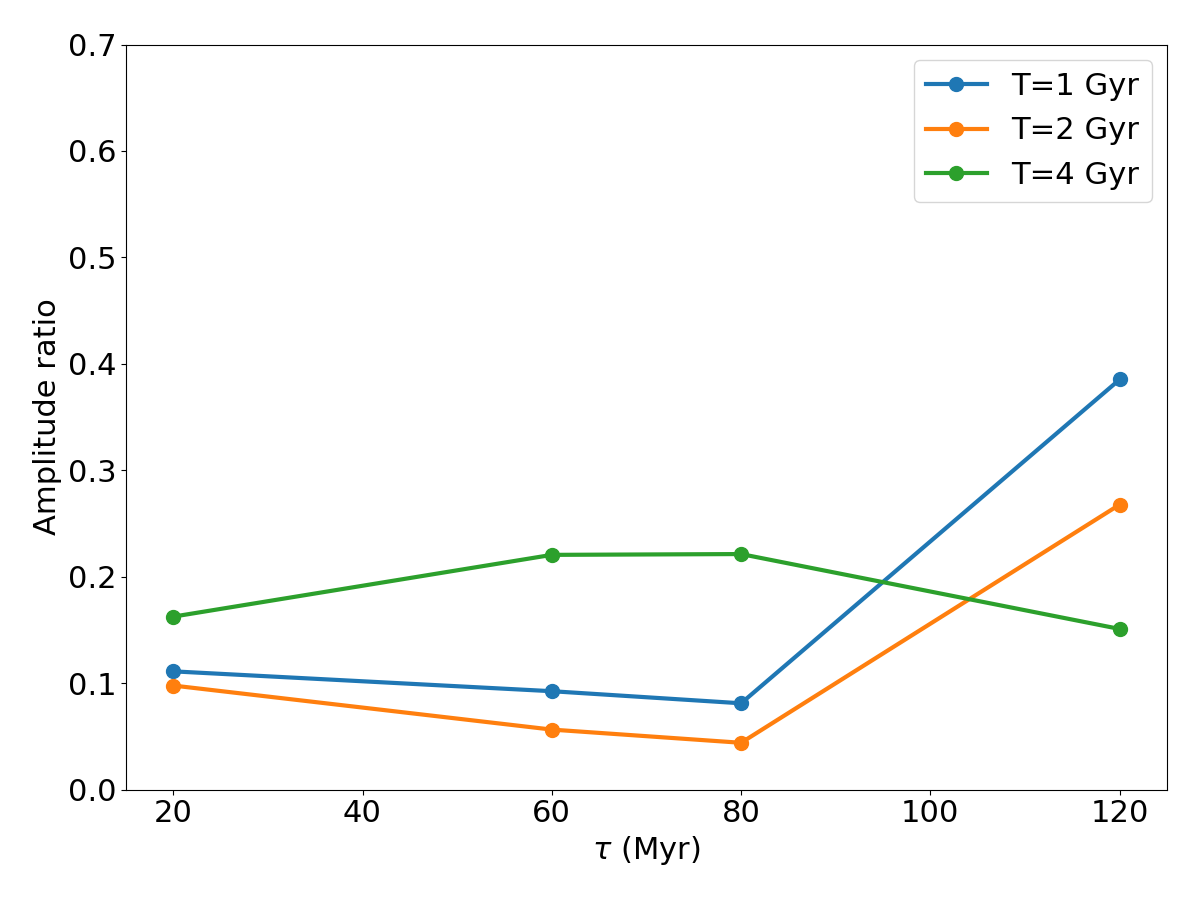}
  \caption{\label{fig:ratios_0p3}As in the top panel of
    Fig. \ref{fig:ratios} but for \(\epsilon=0.3\). Here, the
    suppression ratios are smaller overall with less variation in
    \(\tau_1\).}
\end{figure}

We chose the value of \(\epsilon=0.1\) used in Section \ref{sec:tau}
to be close to the critical threshold for significant suppression of
the bar formation.  For the larger value of \(\epsilon=0.3\), the
amplitude ratios for various values of \(\tau_1\) are shown in Figure
\ref{fig:ratios_0p3}.  The overall values of the suppression ratios
are smaller, as expected, and the sensitivity to value of \(\tau_1\)
are much weaker.

\subsubsection{The effect of feedback on a preexisting bar}
\label{sec:destroy}

\begin{figure}
  \centering
  \includegraphics[width=0.45\textwidth]{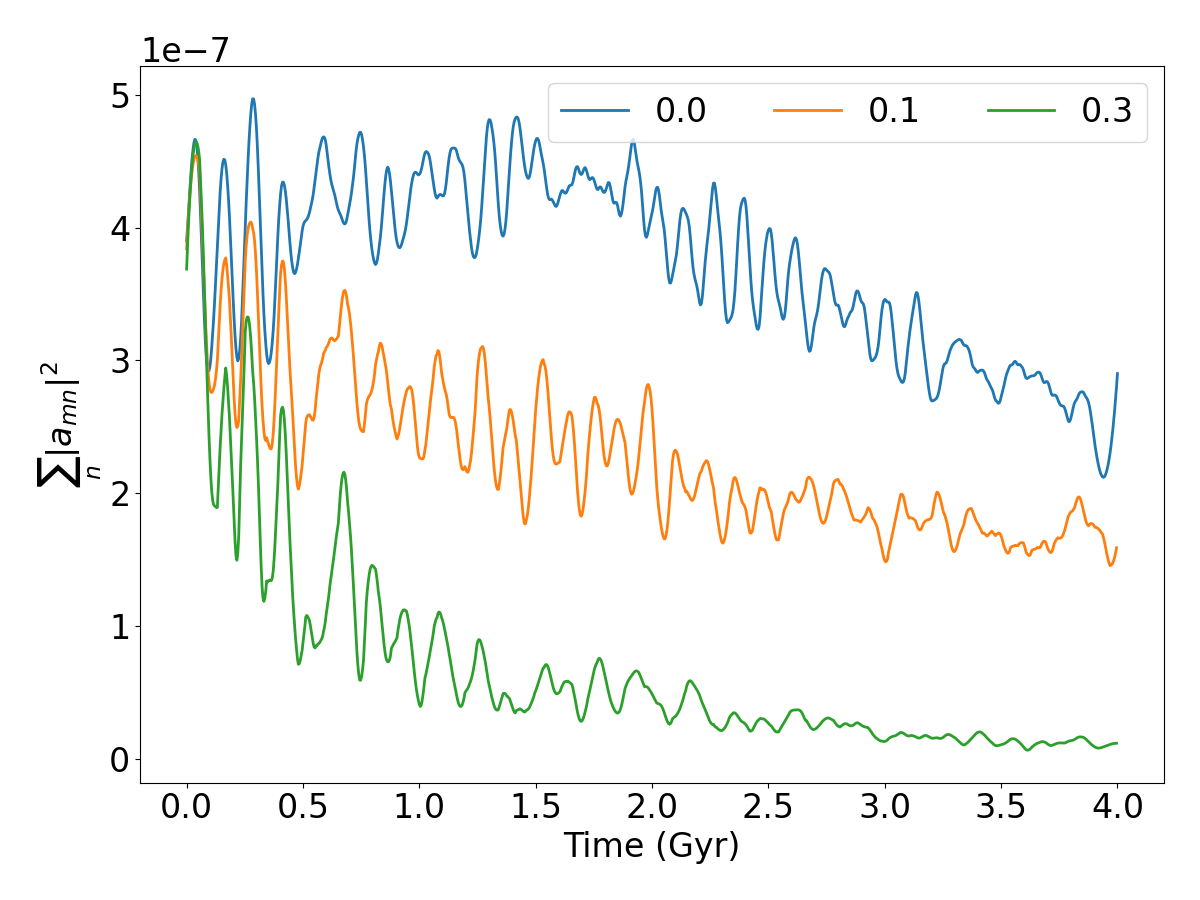}
  \caption{\label{fig:destroy}The primary model is evolved without AGN
    feedback until \(T=2\Gyr\) and then subjected to feedback.  This
    figure shows the \(m=2\) power after feedback commences for
    various amplitudes, \(\epsilon\).  The run with \(\epsilon=0.0\)
    has no feedback, for comparison. A model with \(\epsilon=0.1\)
    reduces the bar amplitude by a factor of 2 and the model with
    \(\epsilon=0.3\) eliminates the bar after 2 Gyr.}
\end{figure}

Bar formation in an otherwise quiescent galaxy could be the source of
gas advection that induces a central star burst.  We ask whether our
Markov-process AGN feedback model can be used to destroy a preexisting
bar?  We begin with the phase space from the primary N-body simulation
without AGN feedback at \(T=2\Gyr\) shown in Figure \ref{fig:densfid}.
We then apply the stochastic model with two different feedback
strengths, \(\epsilon=0.1, 0.3\).  We set \(\tau_1=20\Myr\) with
\(\tau_2=\tau_1/2\) as in previous cases.  These two cases are
compared with the no feedback (\(\epsilon=0\)) case in Figure
\ref{fig:destroy}.  For \(\epsilon=0.1\), the feedback diminishes the
bar amplitude by a factor of 2; the bar reaches a new steady-state
amplitude in approximately 4 Gyr.  For \(\epsilon=0.3\), the bar is
destroyed after 2 Gyr. Some \(m=2\) power remains in the form of an
oval distortion at larger radii than the original bar.

In summary, the AGN feedback model can be used to destroy an existing
bar. More feedback amplitude is required to destroy it than to prevent
it owing to the extra self-gravity in the bar itself.  We suggest
exploring this scenario using existing suites of cosmological galaxy
formation simulations that include feedback.  Barred galaxies are
ubiquitous in simulations of isolated galaxies at the present epoch
although it is clear observationally that not all disk galaxies have
bars.  It has been argued that central mass concentrations may be
sufficient to destroy bars although there is some numerical evidence to
the contrary \citep{Athanassoula:2005}.  The feedback model proposed
here provides another possible bar destruction mechanism.

\subsubsection{Insight from perturbation theory}
\label{sec:barhmf}

Section \ref{sec:pertsim} and Appendix \ref{sec:Hpert} describes a
simulation method tailored to the dynamics of Lynden-Bell's mechanism
specifically.  In essence, each simulation particle contributes its
part to the ensemble action-angle expansion of gravitational potential
restricted to the same commensurate term considered by
\citet{Lynden-Bell:1979}.  Given that this model is a cousin of the
classical cosine HMF ring model \citep{Antoni.Ruffo:1995} specialized
to the interaction implied by the Lynden-Bell mechanism, we find the
term BarHMF to be an apt descriptor.  Although the model employs an
interaction potential from linear theory, it is a non-linear model.
Indeed, Appendix \ref{sec:Hpert} demonstrates that many of the basic
results of bar formation and growth from full N-body simulations are
found in the BarHMF model.  For example, the BarHMF model reproduces
the non-linear saturation of the bar amplitude with a growth rate
proportional to disk mass.

The perturbation scheme outlined in Section \ref{sec:pert} is applied
to the BarHMF simulation as follows.  The axisymmetric component of
the gravitational potential is represented by a spherical monopole for
computational convenience. In other words, the disk does not have its
own \(m=0\) self gravity but does have \(m=2\) self gravity.  The
Markov-process AGN feedback model changes the central mass through
equation (\ref{eq:Pevent}) and causes fluctuations in the monopole
that affect the BarHMF dynamics.  A feedback event implies in an
overall decrease in central mass according to Step 2 in the algorithm
from Section \ref{sec:pert}.  This mass change is applied suddenly as
in the full N-body simulation.  This increases the energy of the
BarHMF particles by \(\delta E = GM_d\epsilon/(\mu\lambda a)\) where
\(\mu\) parametrizes the change in gravitational potential caused by
the feedback-induced mass loss.  We choose \(r_g\equiv\mu\lambda a\)
to be the gravitational radius of the monopole contribution of the
total gravitational energy from change in disk mass inside of
\(\lambda a\).  For a spherical model, \(r_g\) is approximately
\(0.4\) times the half-mass radius of the disk mass enclosed by
\(\lambda a\) \citep{Binney.Tremaine:2008} and determines \(\mu\).
This approximation most likely underestimates the effect on the disk
for a given \(\epsilon\) but is consistent with the assumptions in our
BarHMF application.

The main difference between the AGN model for BarHMF and the full
\EXP\ model is that both the change in energy and recovery time is
instantaneous for BarHMF and only the change in energy is
instantaneous for the \EXP\ model.  This is consistent with the time
averaging implicit in the BarHMF contribution to the interaction
potential: the decay time \(\tau_2\) from equation (\ref{eq:Mdecay})
is shorter than the orbital time and therefore instantaneous in the
time-averaged Hamiltonian context.

The BarHMF particles interact with each other (see Appendix
\ref{sec:Hpert}).  For particles with the same action values, the
coupling depends on the precession angles of their apsides relative to
the bar position angle in some frame rotating at pattern speed
\(\Omega_p\). The precession frequency is small compared to the mean
orbital frequency for orbits comprising the bar.  For example, the
\emph{perfect} bar supporting trajectory has zero precession frequency
with respect to the bar major axis. For a slightly \emph{imperfect}
bar supporting trajectory, the angle of apsides will precess slowly
relative to the bar pattern.  For this reason, the action conjugate to
the precession angle is called the \emph{slow} action and is
proportional to the angular momentum.  The precession angle itself is
called the \emph{slow} angle.  There is a second degree of freedom
whose frequency is proportional to the orbital frequencies and
therefore faster than the precession frequency.  The BarHMF particles
result from an average over this \emph{fast} degree of freedom leaving
only one active \emph{slow} degree-of-freedom.

Without any external perturbations, the fast action is conserved for
all time.  A mass-loss event causes a sudden change in energy and
radial action.  This fluctuation is axisymmetric, conserving the
angular momentum of stellar trajectories; therefore the \emph{slow}
action remains unchanged.  However, the \emph{fast} action is a linear
combination of the radial action and angular momentum.  The fast
action changes as result of the energy causing a jump in radial
action.  If these jumps in fast action are sufficiently large, a
BarHMF particle which had been librating about the about the bar
position can find itself precessing away from the bar.  This causes
the bar to grow more slowly, decay, or not grow at all.

\begin{figure}
  \centering
  \includegraphics[width=0.5\textwidth]{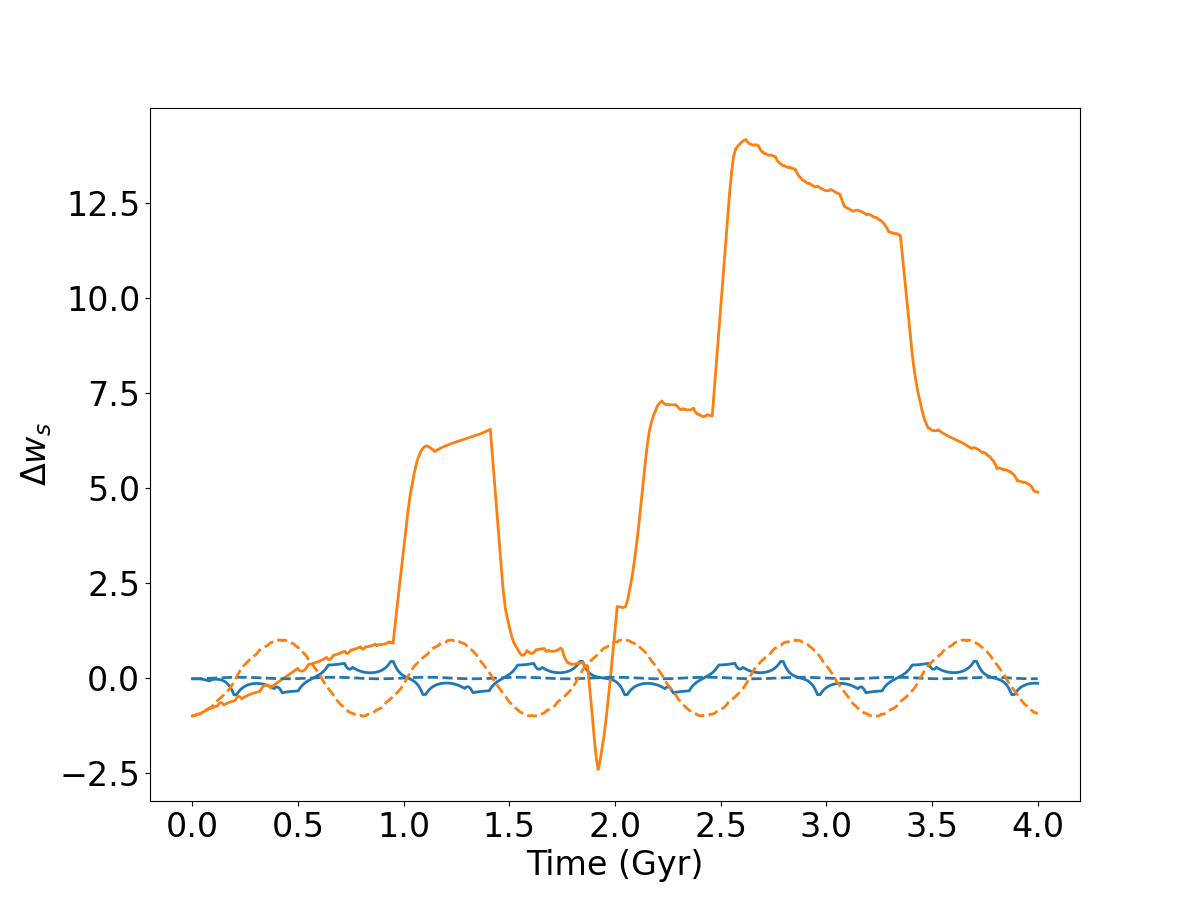}
  \caption{Change in relative precession angle, \(\Delta w_s\), for a
    two-particle BarHMF simulation with AGN feedback.  The mass-loss
    events, which occur every 20 million years push
    \(2\times10^8\msun\) into the inner halo.  We assume that this
    quickly cools back on the disk maintaining a steady state. The
    sudden change in gravitational potential changes the fast action
    for each trajectory.  The two runs differ only in their initial
    \(\Delta w_s\) with 0.02 rad (blue) and 1.0 rad (orange).  The
    dashed curves show the run of \(\Delta w_s\) without feedback.}
  \label{fig:BarHMF_2part_loss}
\end{figure}

We illustrate the dynamics in a simple case of two BarHMF particles at
one disk scale length orbiting in the truncated NFW-like model
(eq. \ref{eq:twopower}).  Each trajectory feels a change in energy
from a mass-loss event every \(\tau_1=20\Myr\).  Each event removes
\(2\times10^8\msun\) from the inner 20\% of the disk scale length
(\(600\pc\) for the Milky Way).  We assume that this quickly cools
back on the disk maintaining a steady state potential punctuated by
instantaneous changes in the radial action.  The mass is not required
to escape the gravitational potential of the galaxy to have affect; it
only needs to pushed out to a few \kpc.  Figure
\ref{fig:BarHMF_2part_loss} shows two BarHMF runs that differ only in
their initial \(\Delta w_s\): the first of these runs begins with
\(\Delta w_s=0.02\) (or \(\approx1^\circ\) shown in blue) and the
second begins with \(\Delta w_s=1.0\) (or \(\approx57^\circ\) shown in
orange).  The first run undergoes significant jumps in \(\Delta w_s\)
but remains in libration.  The jumps in the second with larger
libration amplitude are enough to push the trajectory from libration
into rotation.  The dashed curves depict the same trajectories without
perturbation and can be compared to the two-particle BarHMF runs from
Appendix \ref{sec:Hpert} in Figure \ref{fig:two_body_LB}.

Figures \ref{fig:BarHMF_loss_01} shows the BarHMF model described in
Appendix \ref{sec:Hpert} for the NFW-like background model that is
generally bar promoting.  Each of these simulations has \(N=1000\)
particles\footnote{BarHMF particles are spatially distributed and this
  specialized simulation require fewer particles to reach convergence.
  Conversely, these simulations are more expensive per particle than a
  traditional N-body See Appendix \ref{sec:Hpert} for more
  discussion.}  and uses \(\tau_1=20\Myr\).  The legend describes the
mass loss per AGN event in solar masses scaled to Milky Way units for
familiarity.  For no feedback at all, the bar exponential grows in
approximately 2 Gyr before leveling off into a steady state.  Recall
that there is no coupling between the bar and disk or halo so the bar
can not slow. For \(4\times10^7\msun\) per event, the bar grows more
slowly then levels off and begins to decay.  The overall saturation
amplitude is a factor of 5 smaller than with no feedback at all.  For
a mass loss of \(2\times10^8\msun\) per event, the bar grows but very
slowly, reaching only 10\% of its saturation amplitude in Figure
\ref{fig:powerDemo} at \(T=4\Gyr\).  For larger mass loss events, the
bar can not grow at all.

The thresholds in \(\epsilon\) appear to be larger here than for the
full simulation.  This may have a number of causes.  First, this may
be an artifact of the spherical background model and the gravitational
radius approximation. The work done on the spherical potential from
the feedback is likely to be relatively smaller in effect for the
spherical monopole than for the flat axisymmetric disk. Secondly, the
initial actions are chosen for a fixed eccentricity with guiding
centers in a narrow ring as described in Appendix \ref{sec:Hpert}.
The resulting dynamics will differ from the full simulation that has a
disk with an astronomically motivated exponential distribution.  We
have chosen the total mass of the BarHMF particles to approximate the
mass of the exponential disk between pericenter and apocenter, but we
do not expect a one-to-one correspondence with full N-body
simulations.  Nonetheless, the qualitative trends are similar which
suggest that the BarHMF dynamics represents the main features of the
underlying mechanism.

\begin{figure}
  \includegraphics[width=0.5\textwidth]{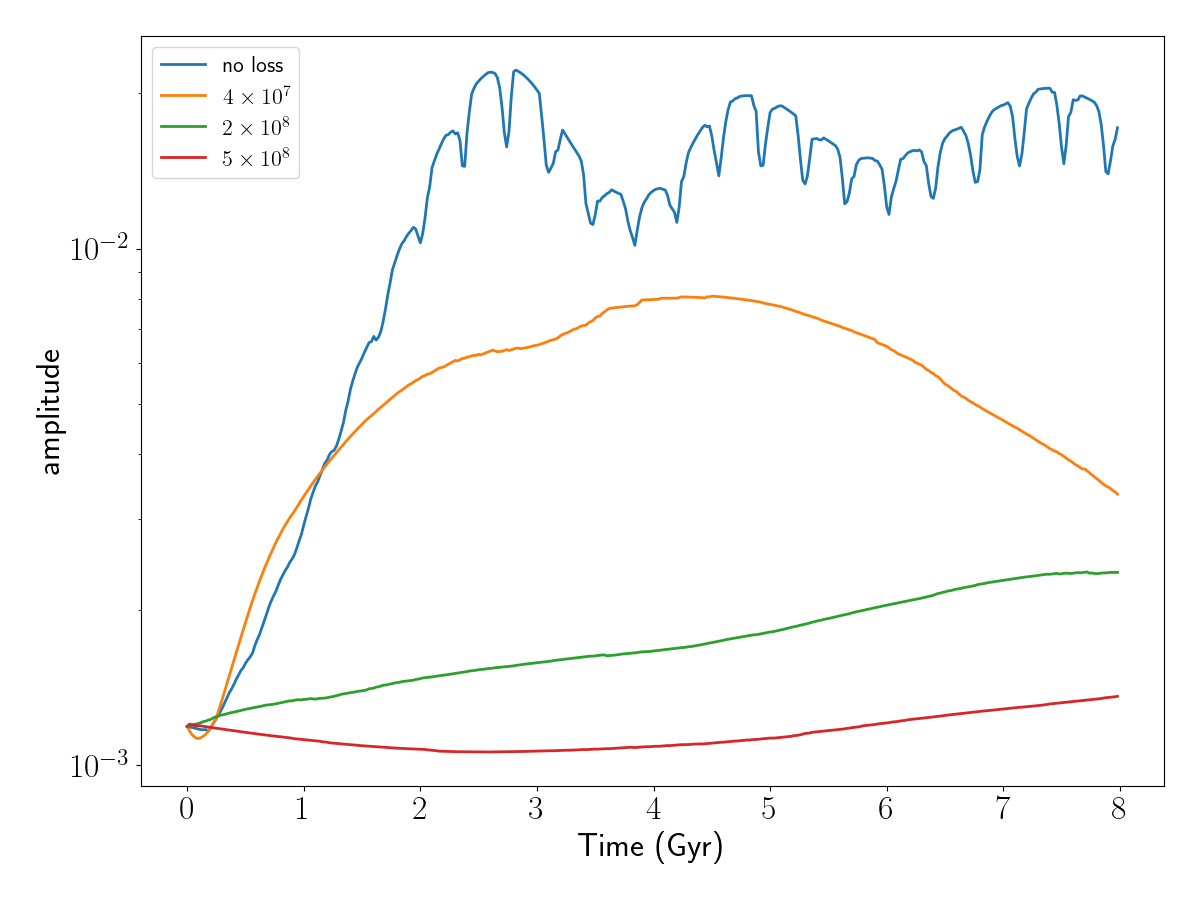}
  \caption{Bar growth as a function of mass loss per event (in
    \(\msun\)) for the BarHMF in Milky Way units of \(10^{12}\msun\)
    for easy comparison.  The event frequency is \(\tau=20\Myr\).}
  \label{fig:BarHMF_loss_01}
\end{figure}

\bigskip\bigskip

\section{Conclusions}
\label{sec:conclu}

We investigate the dynamical importance of feedback-induced
gravitational fluctuations on the bar instability using a Markov
jump-process model; intervals between feedback events are random draws
from a Poisson distribution. We began with a full N-body simulation of
a galactic disk in a live dark-matter halo with a mass typical of a
disk at \(z\sim2.5\) which is roughly a factor of four lower than at
\(z=0\).  Our primary, unperturbed model forms a stable bar in
approximately 1 Gyr.  We model the gravitational influence of AGN
feedback by perturbing that simulation with a stochastic mass-loss
process which removes a small fraction \(\epsilon\) of mass in the
inner disk at a characteristic time \(\tau_1\). The mass is lost from
the inner 2/3 of the disk scale length or 2 kpc in Milky-Way units,
causing a gravitational fluctuation in energy.  The lost mass is
assumed to resettle on the disk at another characteristic time
\(\tau_2\).  We choose \(\tau_2\lesssim\tau_1\) typical of the AGN
event frequency: between 10 and 100 Myr.  We parametrize the mass-loss
strength as the fraction of the disk mass in the inner 2/3 of the disk
scale length in the resulting outflow. The amplitude of the mass loss
ranges from 1\% to 7\% of the total disk mass. The process itself is
designed to capture the temporal features of gravitational
fluctuations from feedback; we intend for ranges in amplitude,
\(\epsilon\), and time scales, \(\tau_1\) and \(\tau_2\),
(Fig. \ref{fig:ratios}) to be compared to detailed physical models.

We also explored a star-formation (SF) model inspired by the
super-bubble model from \citet{Kipper.etal:25}.  We find that the
stochastic feedback from SF outflows in super bubbles does not provide
sufficient correlated dynamical fluctuations to couple to the bar
formation mechanism.  However, this model does \emph{not} include the
work done by a galactic fountain on the potential.  The two models
together suggest that approximately 1\% disk-mass fluctuations in the
inner galaxy are key to coupling feedback to stellar bar suppression.
Specifically, the AGN feedback model also applies to strong SF-driven
outflows.

To gain physical insight into the dynamical nature of the coupling
between feedback and the bar instability, we developed a novel
one-degree-of-freedom simulation based on the classical bar
instability from linear perturbation theory \citep{Lynden-Bell:1979}.
Each particle in the simulation interacts with the bulk of particle
ensemble through the quadrupole force exerted near resonance.  In
other words, each particle is dumbbell-shaped in space with two
actions characterizing its radial extend and spread.  The one degree
of freedom is its azimuthal orientation.  This specialized simulation
reproduces all of the major features the bar instability in the full
N-body simulation including non-linear saturation of the exponential
growth (see Appendix \ref{sec:Hpert}).

We may use this model to understand the dynamics of coupling.  In the
absence of feedback, only the net angular momentum of stellar orbits
change through resonant interaction.  A particular linear combination
of the radial action and angular momentum remains invariant.  The bar
begins with the chance positional alignment of some dumbbell
orientations. Their collective gravity cause more dumbbells to align,
driving an exponential instability.  However, the fluctuations in the
gravitational potential induced by star-formation process affects the
previously invariants, by changing the radial action.  These
fluctuations disturb the libration of particles in the quadrupole of
the bulk, slowing or eliminating the bar instability depending on the
strength of the feedback.

A final caveat: please note that this study considers a single,
isolated galaxy model with an exponential disk and a dark-matter halo
(see Section \ref{sec:model} for details).  The disk mass of
\(5\times 10^9 \msun\) and lack of bulge or spheroid component are
chosen to produce a bar within 2 Gyr without feedback.  This is
consistent with our goal of investigating the dynamics that couples
stochastic events to the bar instability. For example, a more
concentrated model might slow bar formation and change the rate of
pattern speed evolution.

The key findings are as follows:
\begin{enumerate}
\item Mass-loss fractions of \(\epsilon\ge0.1\) in the inner few
  kiloparsecs of a disk are enough to dramatically affect the bar
  formation.  These result in bar amplitude that are at most 50\% of
  the unperturbed case.
\item We demonstrate that fluctuations in the gravitational potential
  of the inner galaxy couple to the radial action of orbits in the bar
  instability.  These fluctuations reduce or eliminate the bar
  instability, consistent with the results of the full N-body
  simulation.  This simple model provides a way of predicting the
  importance of AGN feedback to disk instabilities.
\item We tested both AGN-inspired and SF-inspired feedback models. The
  mass loss is central in the AGN model and disk-distributed for
  SF-inspired model.  We find that the central feedback typical of AGN
  mass outflows provides sufficient correlated change in the
  gravitational potential to affect the bar growth mechanism.
  Conversely, the SF model does not provide sufficient temporally
  correlated change to couple.  Strong, central SF-induced
  fluctuations in the field may occur during massive bursts but are
  not typical of those typically observed at lower redshift.
\item The cadence of feedback events matters.  The bar amplitude
  is reduced by an order of magnitude (and some cases suppressed
  altogether) if \(\tau_1\) roughly matches the typical orbital
  period.  This makes good dynamical sense.  If disk orbits respond to
  the self-gravity of a quadrupole by precessing towards the potential
  well of the quadrupole, the bar amplitude will grow.  However, if
  this precession is disrupted by changes in the underlying potential,
  this growth is reduced or eliminated.
\item For large amplitude mass loss events, e.g. closer to our upper
  limit of 7\% mass changes per event, the amplitude of bar growth is
  suppressed independent of \(\tau_1\).
\item The bars that do form in the presence of significant
  feedback-induced fluctuations are poorly organized in a qualitative
  sense.  They tend to shed mass and lose stability as star-formation
  events continue.
\end{enumerate}

Our overall focus is the mechanism that connects the stochastic nature
of galaxy feedback mechanisms with the smooth orbital motion of
classical galaxy dynamics.  As in dynamics generally, the key
important quantities are frequencies.  A stochastic process can be
characterized by its autocorrelation function, and here, the process
is Poisson, with a characteristic frequency \(\sim1/\tau_1\).  The
coincidence \(\tau_1\sim\tau_{dyn}\) suggests the possibility of
interesting coupling.  The importance of this coupling to the overall
dynamics of the bar instability depends on the details and requires
explicit calculation.  To that end, we provide evidence for its
magnitude and importance by a combination of direct N-body simulation
and a idealized simulation restricted to the dynamics of bar
instability specifically.  The similarity of the behavior in the two
calculations suggests that we have successfully described the
mechanism.

We emphasize that the details of the feedback process have been
greatly simplified in this study.  For example, we have assumed pure
axisymmetric central fluctuations in our AGN feedback model, ignoring
non-axisymmetric dependence.  Rather, we explored the two important
parameters of the simplest possible stochastic model that excites
gravitational fluctuations--an amplitude and characteristic
frequency--and provide predictions for range of those quantities
necessary to have impact on the bar formation.  Motivated by rough
estimates from more detailed observational and theoretical studies, we
find that range of these parameters overlap predicted ranges of
AGN-driven outflow and recurrence.  The expected diversity of galaxy
formation conditions and SMBH masses suggest that the coupling between
feedback and bar formation will not be binary; some systems may still
be able to easily form bars at early epochs and others not.  Moreover,
we have demonstrated that strong feedback coupling may be capable of
eroding and even destroying a bar after formation.

\acknowledgments

MDW thanks Julien Devriendt and Christophe Pichon whose enthusiasm
motivated me to attack this problem.  MDW is grateful to Mike Petersen
and Chris Hamilton for discussions and to Carrie Filion, Chris
Hamilton, Mike Petersen and Christophe Pichon for helpful comments on
the manuscript.  We also thank the anonymous referees who provided
helpful criticism that helped tighten the conclusions. This research
was supported in part by grant NSF PHY-1748958 to the Kavli Institute
for Theoretical Physics (KITP).

%

\begin{appendix}

\label{sec:appendix}

\section{Two-dimensional disk bases}
\label{sec:2ddisk}

The latest version of \EXP\ \citep{EXP} includes two-dimensional
cylindrical disk bases with three-dimensional gravitational fields.
This allows for gravitational couples between two- and
three-dimensional components.  These bases are used for the
simulations in Section \ref{sec:results} that include the coupling to
the dark-matter halo.  For completeness, this section describes the
construction and computation of the two-dimensional cylindrical bases
as used by \EXP.

We construct two-dimensional polar bases using the same empirical
orthogonal function (EOF) technique used to compute the three
dimensional cylindrical basis \citep{Petersen.etal:2022}.  In essence,
we compute the Gram matrix whose entries are the inner product of an
input basis weighted by the density of the desired equilibrium
profile.  The new basis functions are the input basis weighted by each
of the eigenvectors of the Gram matrix.  The construction provides a
basis \emph{conditioned} by the target equilibrium galaxy profile.
The user has the choice of two possible two-dimensional input bases:
\begin{enumerate}
\item The cylindrical Bessel functions of the first kind,
  \(J_m(\alpha_{m,n}R/R_{max})\) where \(\alpha_{m,n}\) is the
  \(n^{th}\) root of \(J_m\) and \(R_{max}\) is an outer boundary.
  Thus, orthogonality is defined over a finite domain by construction.
\item The Clutton-Brock two-dimensional basis \citep{Clutton-Brock:72}
  which orthogonal over the semi-infinite domain.
\end{enumerate}
The user then chooses a conditioning density.  Currently, these are
the exponential disk, the Kuzmin (Toomre) disk, the finite Mestel disk
\citep{Binney.Tremaine:2008}, the tapered Mestel or Zang disk
\citep{Zang:1976}.  The code was written to allow for a user-defined
density function and this will be available in a future release of
\EXP.

The coupling between the two-dimensional disk and other
three-dimensional components (e.g. a dark-matter halo or a bulge)
requires evaluating the gravitational potential and its gradient
everywhere in space.  The three-dimensional gravitational potential is
not \emph{automatically} computed by the two-dimensional recursion
relations that define the biorthogonal functions but can be evaluated
by Hankel transform.  Numerically, this can be done very accurately
over a finite domain with the quasi-discrete Hankel transformation
\citep[QDHT,][]{LiYu:98,Guizar:04}.  The QDHT algorithm exploits
special properties of the Bessel functions and is best suited for
finite density profiles whose two-dimensional gravitational potential
are well-described using the cylindrical Bessel basis.  This is
automatically true for the EOF bases derived from the cylindrical
Bessel functions.  For this reason, the Clutton-Brock basis should
only be used for EOF basis construction for applications where the
three-dimensional gravitational potential is not needed.

The in-plane density, potential and force fields are one-dimensional
functions for each azimuthal order \(m\) and radial order \(n\).
\EXP\ tables the EOF basis solutions on a fine grid and evaluates
these with linear interpolation.  This is numerically efficient and
accurate.  The potential and force fields for \(|z|>0\) require
two-dimensional tables. Similarly, \EXP\ finely-grids these functions
in two-dimensional tables and evaluates them with bilinear
interpolation.

The remainder of this section describes the implementation of the
two-dimensional basis.  We begin in Section \ref{sec:QDHT} with a
summary of the QDHT algorithm followed by its application to the
Hankel transform that defines two-dimensional biorthogonal basis
functions in Section \ref{sec:biorth2d}.  The complete algorithm is
presented in Section \ref{sec:bessalg} followed by a discussion of the
pros and cons of this approach in Section \ref{sec:bessdisc}.

\subsection{A brief description of the QDHT algorithm}
\label{sec:QDHT}

The development here follows \citet{LiYu:98}. The continuous Hankel
transform pair is
\begin{eqnarray}
  g(k) &=& \int_0^\infty dr r f(r) J_{\nu}(kr), \nonumber \\
  f(r) &=& \int_0^\infty dk k g(k) J_{\nu}(kr),
  \label{eq:hankel}
\end{eqnarray}
where \(J_{\nu}\) is the Bessel function of the first kind of integer
order \(\nu\).  The discrete Hankel transform may be obtained from
equation (\ref{eq:hankel}) by imposing
\begin{align}
  g(k) = 0 & \qquad\mbox{for}\ k>K, \nonumber \\
  f(r) = 0 & \qquad\mbox{for}\ r>R. \nonumber
\end{align}
Using this condition in equations (\ref{eq:hankel}) yields:
\begin{eqnarray}
  g(k) &=& \int_0^R dr r f(r) J_{\nu}(kr), \nonumber \\
  f(r) &=& \int_0^K dk k g(k) J_{\nu}(kr).
  \label{eq:dht}
\end{eqnarray}
We can now use the usual Fourier-Bessel expansion \citep{Watson:1941}
to evaluate \(f(r)\) as follows:
\begin{equation}
  f(r) = \sum_{m=1}^\infty C_{\nu m} J_{\nu}\left(a_{\nu m}\frac{r}{R}\right)
  \label{eq:discf}
\end{equation}
for \(0\le r\le R\) where
\begin{equation}
  C_{\nu m} = \frac{2}{R^2 J_{\nu+1}^2(a_{\nu m})} \int_0^R dr r f(r)
    J_{\nu}\left(\alpha_{\nu m}\frac{r}{R}\right)
    \label{eq:discC}
\end{equation}
and \(\alpha_{\nu m}\) is the \(m^{\mbox{th}}\) root of \(J_\nu\).
Equation (\ref{eq:discC}) is a direct application of the well-known
orthogonality relation for Bessel functions of the first kind.  By
inspection, it is clear that:
\begin{equation}
  C_{\nu m} = \frac{2}{R^2J_{\nu+1}^2(a_{\nu m})}
  g\left(\frac{\alpha_{\nu m}}{R}\right).
  \label{eq:Cnu}
\end{equation}

The continuous transform in equations (\ref{eq:dht}) becomes a
discrete transform by choosing the evaluation points for \(r\) and
\(k\) as:
\begin{align}
  r_j &= \frac{\alpha_{\nu j}}{K} \nonumber \\
  k_j &= \frac{\alpha_{\nu j}}{R}.
        \label{eq:knots}
\end{align}
Substituting into equation (\ref{eq:discf}) and using equation
(\ref{eq:Cnu}), we get
\begin{equation}
  f\left(\frac{\alpha_{\nu n}}{V}\right)
  = \sum_{m=1}^\infty \frac{2}{R^2J_{\nu+1}^2(a_{\nu m})}
  g\left(\frac{\alpha_{\nu m}}{R}\right)
  J_{\nu}\left(\frac{\alpha_{\nu m}\alpha_{\nu n}}{RK}\right)
    \label{eq:discf2}
\end{equation}
and by symmetry
\begin{equation}
  g\left(\frac{\alpha_{\nu m}}{R}\right)
  = \sum_{n=1}^\infty \frac{2}{K^2J_{\nu+1}^2(a_{\nu n})}
  f\left(\frac{\alpha_{\nu n}}{K}\right)
  J_{\nu}\left(\frac{\alpha_{\nu m}\alpha_{\nu n}}{RK}\right).
    \label{eq:discg2}
\end{equation}
Truncating the upper limit of sum of the indices \(m\) and \(n\) to
some value \(N\) in equations (\ref{eq:discf2}) and (\ref{eq:discg2})
yields the discrete Hankel transform.  The symmetry in the these two
equations can be made manifest by defining further the two vectors
with elements
\begin{align}
  F_i &= \frac{f(\alpha_{\nu i}/K) R}{J_{\nu+1}(\alpha_{\nu i})}
        \nonumber \\
  G_i &= \frac{g(\alpha_{\nu i}/R) K}{J_{\nu+1}(\alpha_{\nu i})}
        \label{eq:vecFG}
\end{align}
and the matrix with elements
\begin{equation}
  T_{ij} =  \frac{2}{R K J_{\nu+1}(a_{\nu i})  J_{\nu+1}(a_{\nu j})}
  J_{\nu}\left(\frac{\alpha_{\nu m}\alpha_{\nu n}}{RK}\right).
  \label{eq:matT}
\end{equation}
Then, the discrete Hankel transform may be written in matrix notation
as
\begin{align}
  \mathbf{G} &= \mathbf{T}\cdot\mathbf{F}, \nonumber \\
  \mathbf{F} &= \mathbf{T}\cdot\mathbf{G}.
\end{align}
The transform is very accurate numerically, typically one part in
\(10^8\) in algorithm tests, when the target function is
well-represented by a Bessel expansion,

\subsection{Construction of a biorthogonal basis}
\label{sec:biorth2d}

Armed with the finite Hankel transform and an algorithm for its
evaluation, we derive a two-dimensional biorthogonal basis using the
eigenfunctions of the Laplacian described in \citet{Toomre:1963} and
clearly summarized in \S2.6.2 of \citet{Binney.Tremaine:2008}.
Specifically, gravitational potential functions in the form
\begin{equation}
  \Phi_m(r, \phi; k) = e^{im\phi} J_m(kr)
\end{equation}
have the surface density
\begin{equation}
  \Sigma_m(r, \phi; k) = \frac{k}{2\pi G} e^{im\phi} J_m(kr)
\end{equation}
Using the notation from the previous section, let
\begin{align}
  \Phi_{mj}(r, \phi) &= D_{mj} e^{im\phi} J_m(\alpha_{mj}r/R),
                       \nonumber \\
  \Sigma_{mj}(r, \phi) &= (-1)^m\frac{\alpha_{mj}}{2\pi R G} D_{mj} e^{im\phi}
                         J_m(\alpha_{mj}r/R),
  \label{eq:bessbasis}
\end{align}
where \(D_{mj}\) is a normalization.  Using the standard properties of
Bessel functions, we may write:
\begin{align}
  \left(\Phi_{m^\prime j}, \Sigma_{mk}\right)
  &= \int_0^{2\pi} d\phi \int_0^R dr r
    \Phi_{m^\prime j}^\ast(r) \Sigma_{mk}(r) \nonumber \\
  & = D_{mj} D_{mk} \frac{\alpha_{mj}}{R G} \int_0^R dr r
    J_{m}(\alpha_{mj} r/R) J_{m}(\alpha_{mk} r/R) \nonumber \\
  &= \frac{\alpha_{mj}}{2 R G} J_m^2(\alpha_{m j+1}) D_{mj}^2
    \delta_{m^\prime m}  \delta_{jk}.
    \label{eq:norm}
\end{align}
Setting equation (\ref{eq:norm}) equal to one defines the normalizing
coefficient \(D_{mj}\) in equations (\ref{eq:bessbasis}).  Finally, we
can use the EOF conditioning procedure described above and in
\citet{Petersen.etal:2022} to obtain a new biorthogonal basis that
best fits a desired equilibrium density model.  This new basis is
related to equations (\ref{eq:bessbasis}) by a linear transformation.

In the application needed here, we want to describe the gravitational
potential for three dimensional space.  We may use \citet{Toomre:1963}
to do this.  For razor-thin disk with surface density
\(\Sigma(r, \phi)\), the gravitational potential is:
\begin{equation}
  \Phi(r, \phi, z) = \sum_{m=-\infty}^{\infty}
  \int_0^\infty dk S_m(k) J_m(kr)e^{im\phi - k|z|}
\end{equation}
where
\begin{equation}
  \Sigma_m(r) \equiv \frac{1}{2\pi}\int_0^{2\pi} d\phi e^{-im\phi}
  \Sigma(r, \phi)
\end{equation}
and
\begin{equation}
  S_m(k) = -2\pi G\int_0^\infty dr r J_m(kr)\Sigma_m(r)
\end{equation}
If we use a surface density based on the Bessel basis from equations
(\ref{eq:bessbasis}), the integrals are accurately computed from the
discrete Hankel transform described in the previous section.

\begin{figure}
  \centering
  \includegraphics[width=0.85\textwidth]{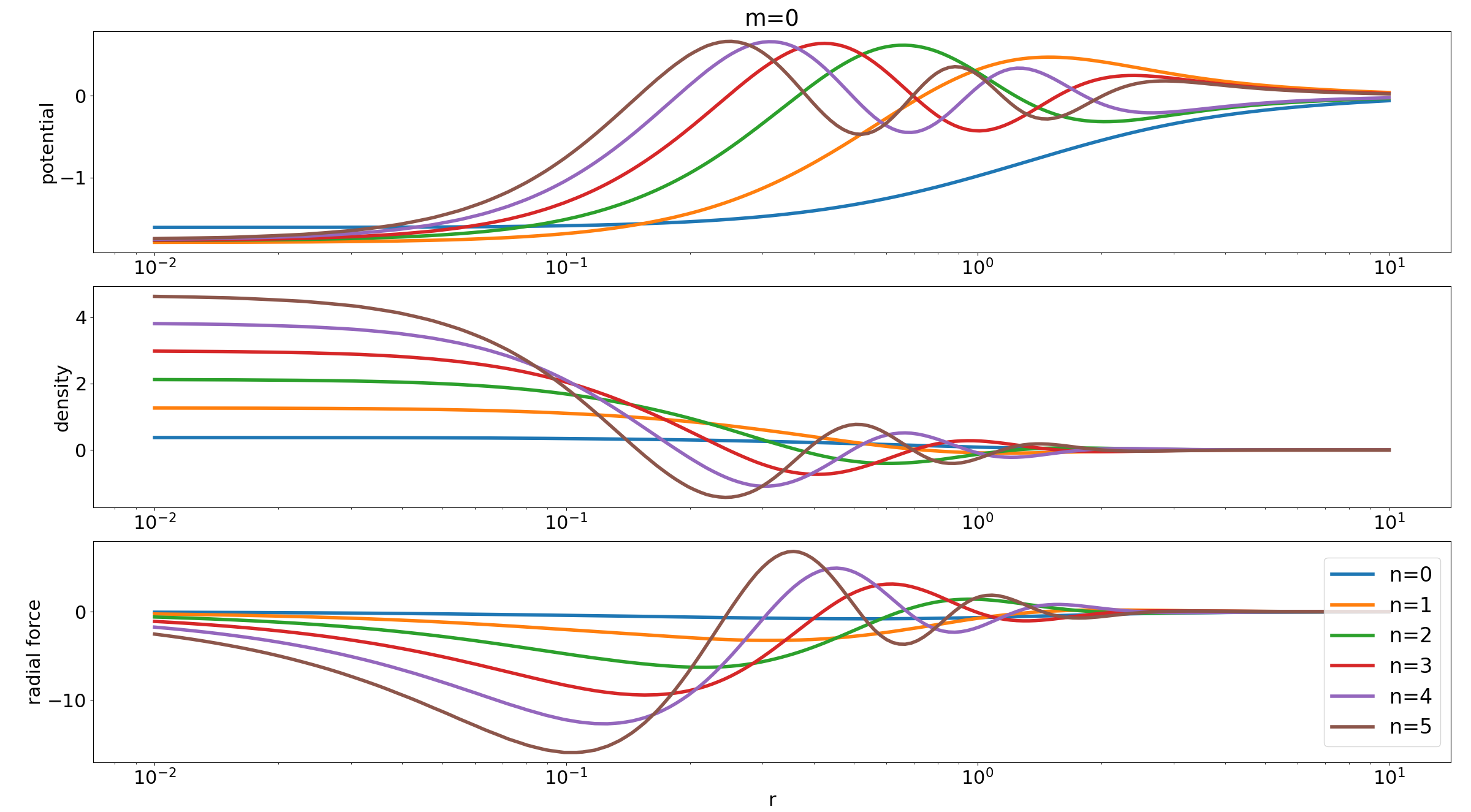}
  \caption{Biorthogonal \(m=0\) basis functions for the
    two-dimensional exponential disk with scale length \(a=1\).
    The color key describes the radial index, \(n\).  The \(n=0\)
    basis function has no radial nodes.  The \(n=1\) function has one
    radial node.  Each successive function has one more radial node.
    The exponential density conditioning leads to nodes distributed
    like the cumulative mass by construction.}
  \label{fig:expon_ortho}
  \vspace{0.45cm} 
\end{figure}

\subsection{Algorithm}
\label{sec:bessalg}

Putting this development together, we may summarize the computation of
the gravitational potential in \(R, \phi, z\) as follows:
\begin{enumerate}
\item Choose the edge of the radial domain: \(R\).
\item Choose a Bessel expansion order: \(N\).
\item Choose \(RK = \alpha_{m,N+1}\) to minimize the error matrix as
  described in \citet{LiYu:98}.  This defines the value
  \(K=\alpha_{m,N+1}/R\).
\item The evaluation knots are then as in equation (\ref{eq:knots})
  for \(j\in[1,\ldots,N]\).
\item Compute the matrix \(\mathbf{T}\) from equation (\ref{eq:matT}).
\item Compute the vector \(\mathbf{F}\) for the surface density
  \(\Sigma_m\) to get the discrete Hankel transform:
  \(\mathbf{G} = \mathbf{T}\cdot\mathbf{F}\).
\item Compute the potential integrand:
  \[
    \mathbf{K} =
    \left\{\frac{e^{-|\mathbf{k}|z}}{\mathbf{k}}\right\}\otimes\mathbf{G}
  \]
  or equivalently \(K_j = -\exp(-k_j|z|) G_j/k_j\).
\item Compute the inverse discrete Hankel transform to get:
  \[\mathbf{\Phi} = \mathbf{T}\cdot\mathbf{K}\].
\item Finally, the evaluation of \(\Phi\) at an arbitrary value of
  \(r\) may be accurately performed using equation (\ref{eq:discg2})
  as an interpolation formula.
\end{enumerate}

\subsection{Discussion}
\label{sec:bessdisc}

The main advantages of the strategy outlined in this section for
numerically evaluating the three-dimensional gravitational potential
for a razor-thin disk are:
\begin{enumerate}
\item One can condition the biorthogonal Bessel-function basis on any
  well-behaved target density, similar to the three-dimensional case.
\item The Hankel transform necessary to obtain the off-plane potential
  is very accurate.
\end{enumerate}

The disadvantages of this approach are:
\begin{enumerate}
\item It works well \emph{only} for a basis with a finite extent.  The
  Hankel transformation for an infinite domain is numerically
  challenging. We have had good success using QDHT, so we recommend
  the finite basis for this reason.  This restriction to a
  finite-domain is not a problem in practice for simulations whose
  implementations are numerically finite.
\item The Bessel functions are asymptotically oscillatory and the
  truncation to order \(N\) will leave their oscillatory imprint at
  larger radii.  Again, this is not a problem for simulations where
  the larger disk radii are not dynamically important.
\end{enumerate}

While the code implemented in \EXP\ allows the conditioning on the 2d
Clutton-Brock basis, the computation of the vertical potential is less
accurate for the same computational work using QDHT. This could be
improved by truncating the basis and implementing Neumann boundary
conditions with appropriate outgoing eigenfunctions.  That has not
been done here.  Another solution might be an asymptotic numerical
evaluation of the Hankel transform at large cylindrical radii patched
onto a brute force quadrature at small radii.  At this point, we
recommend the Bessel basis for accuracy, especially for coupling the
disk response another three-dimensional component such as a bulge or
halo.  If only the two-dimensional in-plane evaluations are needed
(e.g. a two-dimensional disk in a fixed halo), both basis choices will
be equally good.

\section{Hamiltonian perturbation theory}
\label{sec:Hpert}

The bar growth and instability in a two-dimensional disk is mediated
by the inner Lindblad resonance (ILR) as described in
\citet{Lynden-Bell:1979}.  In essence, Lynden-Bell describes bar
growth by determining the conditions for a test-particle orbit to
precess towards the a quadrupole disturbance presented by the
mean-field of all other orbits in the vicinity of the test particle
(see Section \ref{sec:pertsim} for more discussion).

To help understand the N-body dynamics described in Section
\ref{sec:results}, we use the same Hamiltonian perturbation theory to
derive a constrained N-body simulation that excerpts the ILR
interaction from the perturbation theory and treats it as the
fundamental interaction between phase-averaged particles. The new
simulation method is itself non-linear while reproducing the linear
perturbation theory in the linear limit.  The numerical method
parallels the N-body method described in Section \ref{sec:exp}: it
represents the mean gravitational field of the particle ensemble
interacting near the ILR using a biorthogonal expansion.  This
development generalizes the well-known \emph{Hamiltonian Mean Field}
or cosine model \citep{Chavanis.etal:2005} in philosophy, if not
physical detail.  The cosine model assumes that the potential of each
pairwise interaction of particles \(i, j\) on a ring with position
angle \(\theta\) is \(\cos(\theta_i-\theta_j)\).  The model below is a
cosine model whose coefficients depend on actions that exactly
represent the orbit-averaged contributions to the mean field.  The
pairwise interactions are replaced with the interaction of each
particle with the mean contribution of all particles.  For brevity and
with apologies for a small abuse of terminology, we will call this the
\emph{BarHMF} model.  We will see in Section \ref{sec:hmftests} that
the BarHMF model captures many of the interesting linear and
non-linear aspects of bar growth and instability that we see in the
full N-body simulations.  This correspondence allows us to use this
model to help explain the N-body simulations in Section
\ref{sec:results}.

\subsection{Perturbation theory}

The linear response theory has been described in many places and this
development parallels \citet{Tremaine.Weinberg:84} and
\citet{Weinberg:91}.  Our goal here is to present enough of the
mathematical derivation to motivate the hybrid N-body solution.  We
consider a non-axisymmetric gravitational potential:
\begin{equation}
  U_p(\mathbf{r}, t) = \sum_m U_m(r) e^{i m(\phi - \Omega_{pm} t)}.
  \label{eq:perturbedU}
\end{equation}
with a well-defined pattern speed, \(\Omega_{pm}\), for each azimuthal
harmonic, \(m\).  For comparison with the results in Section
\ref{sec:results}, we restrict ourselves to the single \(m=2\) polar
harmonic.  We assume a regular axisymmetric background potential.  The
background potential is independent of the particle dynamics.  The
orbital dynamics may be fully expressed in action-angle coordinates.
We will denote the angle and action vectors by
\((\mathbf{w}, \mathbf{I})\) where index 1 (2) is the radial
(azimuthal) degree of freedom. To make explicit correspondence with
the flat disk considered in this paper, we have \(I_1=I_r\) (radial
action) and \(I_2=J=L_z\) (angular momentum) with \(l_1=l_r\) and
\(l_2=l_\phi\).  This allows us to expand any function of phase space
as Fourier expansion:
\begin{equation}
  U(\mathbf{w}, \mathbf{I}, t) = \sum_{m}\sum_{l_1^\prime=-\infty}^{\infty} \sum_{l_2^\prime=-m}^{m}
  W_{\mathbf{l}^\prime\,m}(\mathbf{I}) e^{i(\mathbf{l}^\prime\cdot\mathbf{w}-m\Omega_{pm}t)},
  \label{eq:Uexp}
\end{equation}
where \(\mathbf{l} = (l_1, l_2)\) and
\begin{equation}
  W_{\mathbf{l}\,m}(\mathbf{I}) = \frac{1}{(2\pi)^2} \oint dw_1 dw_2
  e^{-i(\mathbf{l}\cdot\mathbf{w}-m\Omega_{pm}t)} U(\mathbf{w}, \mathbf{I}, t).
\end{equation}
If we now specialize this to equation (\ref{eq:perturbedU}) with the
assumption of a fixed axisymmetric background and \(m=2\) only, we
find
\begin{equation}
  W_{\mathbf{l}\,2}(\mathbf{I}) = \delta_{l_2\,2}\frac{1}{2\pi} \oint dw_1
  e^{[-i[l_1 w_1 + l_2 f(w_1)]} U_2(r(\mathbf{w}, \mathbf{I}))
\label{eq:radtrans}
\end{equation}
where \(w_2 = \phi + f(w_1)\).  The function \(f(w_1)\) represents the
shift of azimuth from that of the guiding center owing to the radial
motion \citep{Tremaine.Weinberg:84}.  This is a direct consequence of
conservation of angular momentum.  Equation (\ref{eq:radtrans}) is
one-dimensional periodic quadrature that can be numerically computed
to very high accuracy with little overhead.

The radial function \(U_2(r)\) in equation (\ref{eq:perturbedU}) is
arbitrary. Next, we will assume that \(U_2(r)\) can be expanded in the
biorthogonal basis as described in Appendix \ref{sec:2ddisk}:
\(\Phi_{mj}\) for \(j\in[0,n_{max})\).  This yields
\begin{equation}
  U_2(r) = \sum_{j=0}^{n_{max}-1} a_j \Phi_{mj}(r)
  \label{eq:radseries}
\end{equation}
where \(a_j\) are expansion coefficients.  We may explicitly compute
and table the angle transforms for each \(j\)
\begin{equation}
  W_{\mathbf{l}\,2}^j = \delta_{l_2\,2}\frac{1}{2\pi} \oint dw_1
  e^{[-i[l_1 w_1 + l_2 f(w_1)]} \Phi_{2\,j}(r(\mathbf{w}, \mathbf{I}))
  \label{eq:Wtrans}
\end{equation}
over a grid in \(I_1, I_2\) for numerical evaluation.

\subsection{Averaging}

We now apply the averaging principle.  The idea is simple: near a
particular resonance, the libration angle defined as
\(\mathbf{l}\cdot\mathbf{w} - \Omega_{p2} t\) is very slowly changing
relative to either component of \(\mathbf{w}\).  This allows us to
make a simple canonical transformation to a linear combination of
angles and actions where one of the new coordinates is this libration
angle by construction.  Let the new angles and actions be \({\cal
  W}=(w_s, w_f)\) and \({\cal I}=(I_s, I_f)\), respectively.  Let us
choose the Type 2 generating function \citep{Goldstein.etal:02}
\begin{equation}
  F_2(\mathbf{\cal I}, \mathbf{w}) = (\mathbf{l}\cdot\mathbf{w} -
  \Omega_{p2} t)I_s + w_1 I_f.
  \label{eq:F2}
\end{equation}
Simple calculation immediately shows that \(w_s =
\mathbf{l}\cdot\mathbf{w} - \Omega_{p2} t\), \(w_f = w_1\), \(I_1 =
l_1 I_s + I_f\) and \(I_2 = l_2 I_s\).  The new Hamiltonian is \({\cal
  H} = H - 2\Omega_{p2} I_s\).  There is some arbitrariness in the
choice of \(F_2\) but the important feature is that \(w_s\) changes
slowly compared to the change in \(w_f\).

Armed with this development, we average equation (\ref{eq:Uexp}) over
a time interval \(T\) chosen so that
\(2\pi/\Omega_1 \ll T \ll 2\pi/(\mathbf{l}\cdot\mathbf{\Omega} -
2\Omega_{p2})\).  By construction, all terms in the sum are rapidly
varying in this interval except for the one with \(l_1^\prime=l_1\)
and \(l_2^\prime=l_2\).  Our perturbed Hamiltonian for the averaged
system reduces to a single term:
\begin{equation}
  H_1(\mathbf{w}, \mathbf{I}) = W_{\mathbf{l}\,2}(\mathbf{I})e^{iw_s}.
\end{equation}
Expressed in the biorthogonal basis, this may be written as
\begin{equation}
  H_1(\mathbf{w}, \mathbf{I}) = \sum_{j=0}^{n_{max}-1} a_j W^j_{\mathbf{l}\,2}(\mathbf{I})e^{iw_s}.
  \label{eq:biorthdef}
\end{equation}

Now, let us take some ensemble of \(N\) particle trajectories.  At any
particular time \(t\), a single trajectory indexed by \(k\) contributes
\begin{eqnarray}
  \delta a_j &=& \int d\mathbf{r} \delta(\mathbf{r} - \mathbf{r}_k(t))
                 \Phi_{mj}(r) e^{-i m(\phi-\Omega_{p2} t)} \nonumber \\
             &=& \int dr \delta(r - r_k(t)) \Phi_{mj}(r) \int
                 d\phi \delta(\phi - \phi_k(t)) e^{-i m(\phi-\Omega_{p2} t)}
\end{eqnarray}
to the coefficient \(a_j\) where \(\delta(\cdot)\) is the Dirac delta
function.  The averaging principle requires an average of the fast
action to get the \emph{slow} contribution.  The contribution of a
single averaged trajectory indexed by \(k\) to \(a_j\) is
\begin{eqnarray}
  a^k_j &=& \frac{1}{2\pi}\oint dw_f \delta a_j \nonumber \\
  &=& \frac{\delta_{l_2, m}}{2\pi}\oint dw_f \Phi_{l_2\,j}(r(\mathbf{w},
  \mathbf{I})) e^{-i l_2(w_2 -  f(w_1) - m\Omega_{p2} t)} \nonumber \\
  &=& e^{-i(l_1 w_1 +  l_2 w_2
    - m\Omega_{p2} t)}
  \frac{\delta_{l_2, m}}{2\pi} \oint dw_1 \Phi_{l_2\,j}(r(\mathbf{w},
  \mathbf{I})) 
  e^{i (l_1 w_1 + l_2 f(w_1))} \nonumber \\
  &=& e^{-iw_s} W_{\mathbf{l}\,2}^{j\ast}(\mathbf{I}),
\end{eqnarray}
where we used \(w_2 = \phi + f(w_1)\) in the second equality,
explicitly held \(w_s\) fixed during the average of \(w_f=w_1\) in the
third equality, and identified \(W_{\mathbf{l}\,2}^j(\mathbf{I})\) from
equation (\ref{eq:Wtrans}). The values of \(w_s\) and \(\mathbf{I}\)
are those of trajectory \(k\).

The Hamilton equations for a particular phase trajectory then
become:
\begin{align}
  \dot{w}_s &=\quad\frac{\partial H}{\partial I_s} = \frac{\partial{\cal
              H}}{\partial I_s} +  \frac{\partial H_1}{\partial I_s}
  =& \mathbf{l}\cdot\mathbf{\Omega} - 2\Omega_{p2} +
    e^{iw_s} \sum_{j=0}^{n_{max}-1} a_j \frac{\partial
    W^j_{\mathbf{l}\,2}(\mathbf{I})}{\partial I_s}, && \label{eq:Hpert1} \\
  \dot{I}_s &= -\frac{\partial H}{\partial w_s}
  =& -ie^{iw_s} \sum_{j=0}^{n_{max}-1} a_j
     W^j_{\mathbf{l}\,2}(\mathbf{I}). &&
    \label{eq:Hpert2}
\end{align}
The first term in equation (\ref{eq:Hpert1}) is the unperturbed
frequency of the slow motion.  This frequency vanishes at the
resonance.  The final terms in each of equations (\ref{eq:Hpert1}) and
(\ref{eq:Hpert2}) describe the perturbation from the quadrupole
component contributed by each particle.  This is a pendulum equation
where the pendulum arm has a general dependence on the action of the
particle.

\subsection{The hybrid simulation method}

We begin by adopting the biorthogonal basis from Appendix
\ref{sec:2ddisk} that matches the simulations in Section
\ref{sec:results}.  We represent the axisymmetric background potential
by the spherical halo model from Section \ref{sec:halo} for
computational convenience and simplicity; using a combined disk,
bulge, dark-matter halo background field is straightforward but not
necessary for this demonstration.  The particle motion remains planar
by construction.  Then, we tabulate the values of
\(W^k_{\mathbf{l}\,2}(\mathbf{I})\) on a two-dimensional grid in
\(\mathbf{I}\).  The actions of the simulation particles can be chosen
according to some distribution.  For studies here, we distribute
guiding centers in a smooth distribution in a particular annulus.  The
initial distribution of \(w_s\) is chosen in one of two ways: (1)
random in \([0, 2\pi]\) or (2) uniformly spaced in \([0, 2\pi)\) with
an specific excess in a small interval of \((-\epsilon, \epsilon)\) to
seed the perturbation.  For Case 1, the small variations from uniform
are sufficient to seed an instability for a bar-unstable system.  For
Case 2 without an excess, a uniformly distribution will remain close
to equilibrium for an unreasonably long time. In other words, this
choice relies on numerical noise to produce an asymmetry that grows.
This motivates adding a small deviation to seed a disturbance.

Our simulation particles only contribute their quadrupole component
from a single \(\mathbf{l}\) term in equation (\ref{eq:Uexp}) to the
gravitational field.  Each particle begins with a mass, actions
\(I_f\) and \(I_s\) and slow angle \(w_s\).  Without external
influences, the fast action \(I_f\) is fixed throughout. However,
under the influence of the Markov jump process described in Section
\ref{sec:pert}, both \(I_f\) and \(I_s\) can change as described in
Section \ref{sec:barhmf}.  The fast action, \(I_f\), changes result
from the gravitational fluctuations and the slow action, \(I_s\),
changes as a result of the quadrupole field.  Assume that we have
\(N\) particles of mass \(m_k\) distributed as described in the
previous paragraph.  The mean-field potential is determined by summing
the contributions of each particle to the \(n_{max}\)
numerically-determined coefficients \(\hat{a}_j\):
\begin{equation}
  \hat{a}_j = \sum_{k=1}^N m_k a^k_j = \sum_{k=1}^N m_k e^{-iw_k}
    W^{j\ast}_{\mathbf{l}\,2}(\mathbf{I}_k).
    \label{eq:coefj}
\end{equation}
The averaging smooths out each particle's spatial contribution from
pericentre to apocentre. This results in a faster convergence with
particle number \(N\) and radial order \(n_{max}\) than a traditional
N-body code.  If we define \(J(E)\) as the maximum value of orbital
angular momentum, \(J\), at particular energy \(E\), a converged
series is obtained for \(n_{max}=6\) for modestly eccentric orbits
with \(J/J(E)=0.8\).

The simulation proceeds by solving \(2N\) first-order ODE defined by
equations (\ref{eq:Hpert1}) and (\ref{eq:Hpert2}) for \(w_s\) and
\(I_s\) respectively.  Without any external perturbations, the fast
actions \(I_f\) are constant for the entire simulation.  However, the
background values for \(\Omega_{1}(\mathbf{I})\),
\(\Omega_{2}(\mathbf{I})\), \(W^k_{\mathbf{l}\,2}(\mathbf{I})\) and other
phase-space quantities needed by equations (\ref{eq:Hpert1}) and
(\ref{eq:Hpert2}) are naturally tabulated in \(I_1\) and \(I_2\) or
\(E\) and \(L=I_2\).  Thus, any changes in \(I_s\) imply both changes
\(E\) and \(L\).  New values are computed using bilinear interpolation
on precomputed grids of size \(400\times400\).

Unfortunately, this Hamiltonian flow is not explicitly separable which
complicates the choice of an ODE solver. Recall that a Hamiltonian
system is \emph{separable} if the Hamiltonian can be written as
\(H(\mathbf{q}, \mathbf{p}) = K(\mathbf{p})+V(\mathbf{q})\) where
\(K\) and \(V\) most often correspond to the kinetic and potential
functions, and is non-separable otherwise.  A separable system can be
solved explicitly, using the well-known leapfrog algorithm or
higher-order generalizations \citep{Yoshida:1990}.  This approach
works because the Hamiltonian can be separated or \emph{split} into
two pieces that depend only on \(\mathbf{q}\) or \(\mathbf{p}\).  For
example, flows with \(H(\mathbf{q}, \mathbf{p}) = K(\mathbf{p})\) and
\(H(\mathbf{q}, \mathbf{p}) = V (\mathbf{q})\) are both exactly
solvable (see \citealt{Sanz-Serna.Calvo:2018} for examples).  To avoid
an implicit method for the non-separable equations
(\ref{eq:Hpert1})--(\ref{eq:Hpert2}), we adopt the clever solution
proposed by \citet{Tao:2016}.  Tao introduces an extended phase space with an
\emph{extra} copy of phase space \((\mathbf{x}, \mathbf{y})\) that
coincides with \((\mathbf{q}, \mathbf{p})\) initially.  The
Hamiltonian of the extended system is the sum of three Hamiltonians:
\begin{equation}
  \bar{H}(\mathbf{q}, \mathbf{p}, \mathbf{x}, \mathbf{y}) =
  H(\mathbf{q}, \mathbf{y}) + H(\mathbf{x}, \mathbf{p}) +
  \frac{\omega}{2}\left[(\mathbf{q} - \mathbf{x})^2 +
    (\mathbf{p}- \mathbf{y})^2\right].
  \label{eq:Tao}
\end{equation}
The first two terms are the original Hamiltonian function and the
third term is a non-physical linear force that drives the two copies
towards each other.  One can solve each of the three Hamiltonians with
an explicit method.  Recall that an explicit sympletic ODE solver is a
Hamiltonian map.  The combined mapping for the three separable
Hamiltonians in equation (\ref{eq:Tao}) is the composition of three
maps.  The coupling parameter \(\omega\) needs to be chosen large
enough to keep the two phase-space copies, \((\mathbf{q},
\mathbf{p})\) and \((\mathbf{x}, \mathbf{y})\), close to each other
but the solution is not sensitive to \(\omega\) once so tuned.  Tao
estimates that the error for a solution of length \(T\) is \({\cal
  O}(Th^l\omega)\) for a duration \({\cal O}(\min[h^{-l}\omega^{-1},
  \omega^{1/2}])\) where \(l\) is the order using
\citet{Yoshida:1990}.  \citet{Tao:2016} gives a detailed discussion of
convergence and consistency. In practice, the Tao method produces
high-accuracy solutions with little energy drift for the conservative
time-independent system.  However, unstable critical systems require
large values of \(\omega\) and some tuning effort especially for
\(N\lesssim400\).  We compared our implementation of Tao's algorithm
to fourth-order Runga Kutta for verification.  In practice, this
method requires a much smaller time step than a traditional N-body
code.  The fourth-order (\(l=4\)) Tao method requires a comparable or
larger number of function evaluations with similar long-term energy
drift as fourth-order Runga Kutta at the same step size.

\subsection{Test results}
\label{sec:hmftests}

The simplest possible test of the new simulation method is the
two-body quadrupole problem: two averaged particles that feel each
other's mutual potential along with the fixed monopole of the
background system.  This system is coupled pendulum and admits
irregular motion.  In the regular domain, the relative motion of the
two interacting particles is oscillatory.  Our first test solves
equations (\ref{eq:Hpert1}) and (\ref{eq:Hpert2}) for \(N=2\) and
plots the the difference in \(w_s\).  The two particles have identical
actions corresponding to a particles with \(J/J_{max}(E)=0.8\) with
guiding center radius of \(r_{c} = 3\kpc\); this corresponds to
approximately one Milky-Way exponential scale length.  The pericenter
and apocenter of the unaveraged orbit is 0.006 and 0.012, respectively
(\(1.8\) and \(3.6\kpc\) in Milky Way units, respectively).  The
initial position angles are displaced by 0.1 radians.  The total mass
of the two-particle system is \(2\times10^{-3}\) of the total
background mass, \(M=2\times10^9\msun\).  These values are chosen to
represent a typical ensemble of disk orbits in the vicinity of the
disk scale length. None of these results depend qualitatively on this
choice.

\begin{figure*}
  \begin{minipage}{0.49\textwidth}
    \includegraphics[width=\textwidth]{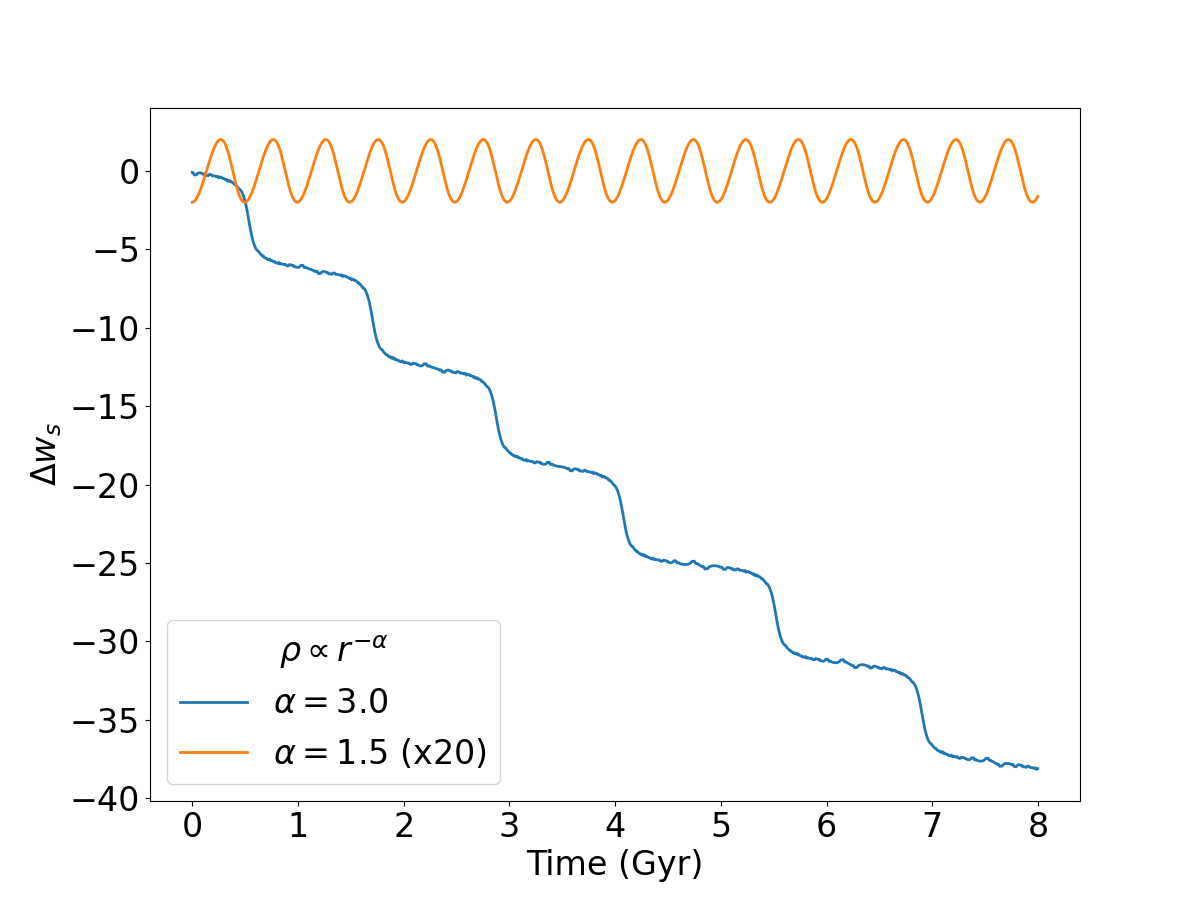}
    \caption{Difference in slow angle, \(\Delta w_s\), for two
      particles with differently sloped rotation curves.  The
      \(\alpha=1.5\) model has a rising rotation curve with a positive
      Lydnen-Bell criterion value \({\cal L}\) and \(\alpha=3.0\) has
      a falling rotation curve with a negative \({\cal L}\).  The
      \(\alpha=1.5\) curve is multiplied by factor of 20 to show the
      oscillations which would be too small to see otherwise.}
    \label{fig:two_body_LB}
  \end{minipage}
  \hspace{0.02\textwidth}
  \begin{minipage}{0.49\textwidth}
  \includegraphics[width=\textwidth]{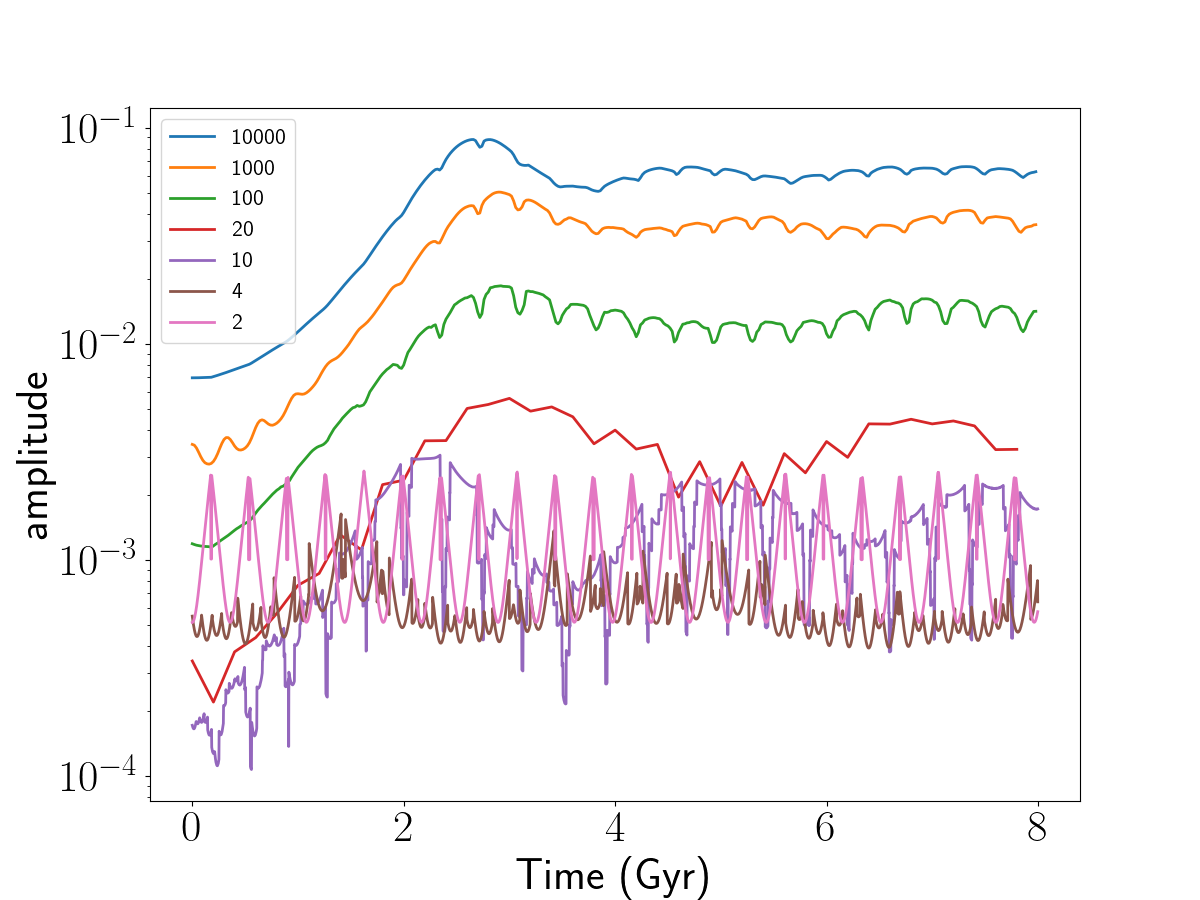}
  \caption{Bar amplitude as a function of particle number \(N\) for
    the NFW-like background model with \(\rho\propto r^{-1}\) in the
    vicinity of the particles.  The amplitude is computed from a
    kernel density estimate of the distribution of \(w_s\) for the
    \(N\) particles.}
  \vspace{2\baselineskip}
  \label{fig:BarHMF_number}
\end{minipage}
\end{figure*}

We consider two different background models (eq. \ref{eq:twopower}):
one with \(\alpha=1.5, \beta=0\) and one \(\alpha=3.0, \beta=0\) with
the default values of \(r_t=1.5, r_w=0.2\) as described in Section
\ref{sec:halo}. The Lynden-Bell criterion from equation (\ref{eq:LB})
predicts that the first should bar \emph{promoting} while the second
should be bar \emph{avoiding}.  That is, the orbits will attract in
the first case or \emph{librate} and repel in the second case or
\emph{rotate}.  Figure \ref{fig:two_body_LB} presents the solutions of
equations (\ref{eq:Hpert1}) and (\ref{eq:Hpert2}) and plots the the
difference in \(w_s\).  This demonstrates the predicted behavior: the
two particles librate for the \(\alpha=1.5\) case which has \({\cal
  L}>0\) and rotate for the \(\alpha=3.0\) case which has \({\cal
  L}<0\).

For \(N\gg1\), the particles can exchange slow action \(I_s\) with the
mean field and thereby with each other.  We use the extremum excursion
of the probability distribution for the \(N\) values of \(w_s\) in the
interval \([0, 2\pi)\) to estimate the \emph{amplitude} of the
  resulting bar.  The distribution is constructed using a kernel
  density estimator with the optimal Gaussian kernel width expression
  from \citet{Silverman:86}.  To explore the change in bar amplitude
  with time for simulations of various \(N\), we choose an the
  NFW-like background model (eq. \ref{eq:twopower} with \(\alpha=1,
  \beta=2\)).  The model has \(\rho\propto r^{-1}\) in the vicinity of
  the BarHMF particles and is explicitly bar promoting according to
  \({\cal L}\).  The particle parameters are the same as the
  two-particle model with the same total particle mass, \(M_b\):
  particles have identical actions corresponding to a particles with
  \(J/J_{max}(E)=0.8\) and a single guiding center radius of \(r_{c} =
  0.01\).

Figure \ref{fig:BarHMF_number} compares the bar amplitude from BarHMF
runs with increasing values of \(N\).  The behavior for \(N=2\) is
sinusoidal libration as in Figure \ref{fig:two_body_LB}.  For
\(N\gtrsim10\), we begin to see a clear period of exponential growth
followed by oscillation.  As \(N\) increases, the amplitude increases
with otherwise similar behavior: the classic bar-like exponential
growth phase followed by a steady state.  The short-period modulation
of \(\approx0.18\) is the one-particle libration time.  This is
corresponds to \(\sim300\Myr\) in Milky Way units.  These modulations
become weaker with increasing \(N\) as expected.  This figure also
shows a longer period modulation in the steady-state phase that
decreases in amplitude and possibly increases in period with
increasing \(N\).  We speculate that this is the ongoing phase mixing
of transient from the initial formation but have not yet investigated
this in detail.

\begin{figure}
  \begin{minipage}{0.49\textwidth}
    \includegraphics[width=\textwidth]{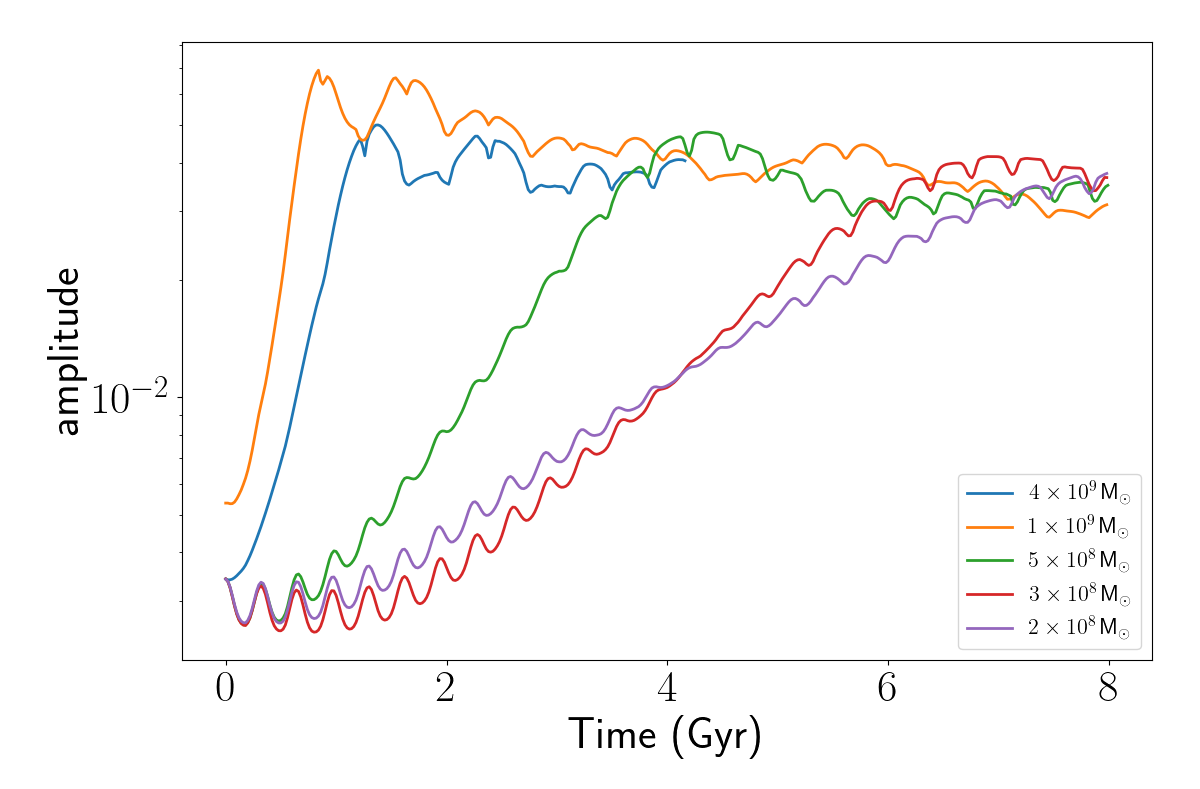}
    \caption{\label{fig:BarHMF_mass} Bar amplitude variation as a
      function of disk mass, showing that the initial exponential
      growth is proportional to bar-particle mass.  The model is the
      same as the one shown in Figure \ref{fig:BarHMF_number} and the
      total particle mass is shown in the legend.}
  \vspace{4\baselineskip}
  \end{minipage}
  \hspace{0.02\textwidth}
  \begin{minipage}{0.49\textwidth}
  \includegraphics[width=\textwidth]{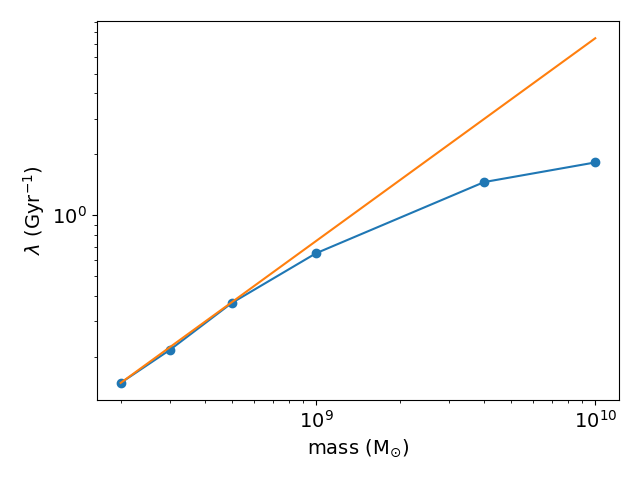}
  \caption{\label{fig:BarHMF_rate} Exponential growth rate for the
    amplitude as a function of mass (blue) estimated from the curves
    in Fig. \ref{fig:BarHMF_mass}.  The expected linear relation
    extrapolated from small mass is shown for comparison (orange).
    The expected linear scaling is only valid for small disk mass.}
  \end{minipage}
\end{figure}

\begin{figure}
  \begin{minipage}{0.49\textwidth}
  \includegraphics[width=\textwidth]{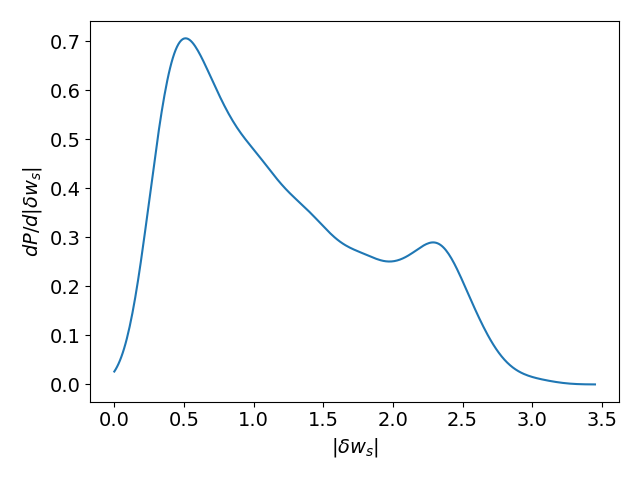}
  \caption{The distribution of the excursion angle, \(|\delta w_s|\),
    in radians for the 9377 librating trajectories for the BarHMF run
    with \(N=10000\) shown in Fig. \ref{fig:BarHMF_number}. The
    excursion in apocentric position is \(|\delta w_s|/2\).  The
    apocentric distribution is approximately flat between \(10^\circ\)
  and \(70^\circ\) with a peak at \(20^\circ\).}
    \label{fig:BarHMF_libration}
  \end{minipage}
\end{figure}

The exponential growth rate seen in Figure \ref{fig:BarHMF_number} is
nearly the same for all \(N\gtrsim100\). Indeed, we expect the growth
rate to depend on and increase linearly with total particle mass,
\(M_b\).  Figure \ref{fig:BarHMF_mass} demonstrates this with a
sequence of simulations with \(N=1000\) and varying mass with \(M_b\)
between \(10^{8}\msun\) and \(10^{10}\msun\).  The growth rate is
estimated as the slope of the rise in the log-linear plot, shown in
Figure \ref{fig:BarHMF_rate}.  This slope is linearly proportional
\(M_b\) until \(M_b\approx10^{9}\msun\).  For larger values, the slope
increases more weakly with mass.  For \(M_b\gtrsim4\times10^{9}\msun\)
the increase in slope is weakly than linear, suggesting that bar is no
longer following the linear predictions.  Rather, it is growing as
fast as it can limited by the intrinsic characteristic time of BarHMF
model.  The regime is typical of bar growth in unstable high-mass
disks typical of N-body simulations of isolated galaxies. 

For a final demonstration of the correspondence between the BarHMF and
the full N-body simulations, we compare the fraction of trapped,
librating orbits after bar formation is complete with the untrapped,
rotating fraction.  We do this by computing the difference between the
position angle of each particle and the mean position angle of the
entire ensemble as a function of time.  The variation of the
distribution around the ring is very close to sinusoidal.  We use the
same the probability distribution estimate of \(w_s\) used to infer
the amplitude to find the position angle of the peak, \(\theta_{PA}\).
This yields a time series of excursions for the angle of each
quadrupole particle relative to the bar position angle: \(\delta
w_s(t) = w_s(t) - \theta_{PA}(t)\).  Then, we estimate the width of
\(\delta w_s\) for each trajectory by computing the distance between
the extremal values for all \(t\), denoted as \(|\delta w_s|\).  A
librating (rotating) trajectory has \(|\delta w_s|\) smaller (larger)
than \(\pi\).  Using this to classify BarHMF trajectories, we find
that 94\% of the distribution is in libration by the end of the
simulation at \(T=4\Gyr\).  Figure \ref{fig:BarHMF_libration} shows
the distribution of widths \(|\delta w_s|\) for the ensemble of
librating trajectories.  This distributions peaks at
\(0.7\,\mbox{rad}\) (or \(40^\circ\)) with a long heavy tail to large
libration angle.  This implies an excursion about the bar position at
apocenter of \(|\delta w_s|/2\) or \(0.35\,\mbox{rad}\) (or
\(20^\circ\)).  \citet{Petersen.etal:2021} find a similarly large
trapped fraction of \(x_1\) orbits of approximately 80\% in the bar
vicinity.  Moreover, the trapped BarHMF shows a broad tail of
libration amplitudes which is also similar to the distributions seen
in N-body simulations.  A precise correspondence to the N-body
simulations from Section \ref{sec:results} should not be expected for
two reasons: (1) we have not included the background potential of the
disk itself; and (2) we have not fully distributed the BarHMF
particles according to the disk phase-space distribution function.
Such simulations are possible but would require an intensive
computational campaign that is beyond the scope of this preliminary
study.  Nonetheless, the good correspondence between the major
features of bar growth and saturation suggest that the simple
perturbation theory captures much of the bar instability mechanism and
supplies a good model for further applications
(Sec. \ref{sec:hmfdiscuss}).

\subsection{Topics for future work}
\label{sec:hmfdiscuss}

This paper focuses on the \emph{classic} \(m=2, l_1=-1, l_2=2\) bar
instability.  Only one term in equations (\ref{eq:perturbedU}) and
(\ref{eq:Uexp}) is required.  However, the BarHMF simulation can be
trivially generalized to include multiple resonances
\((m, l_1, l_2)\), The extended model can be used, for example, to
explore non-linear resonance coupling in barred system in the
non-linear regime.  Technically, this requires adding multiple
commensurabilities of the form in equation (\ref{eq:Uexp}) and
eliminating averaging in favor of the full equations of motion for
\((\mathbf{w}, \mathbf{I})\).  For example, consider adding the
classic 4:1 bar resonance \citep{Contopoulos:1988}.  The pattern speed
is the same for each harmonic:
\(\Omega_p\equiv\Omega_{p2}=\Omega_{p4}\). The commensurability is
\(\Omega_r = 4(\Omega_\phi - \Omega_p)\) and gives an new term with
\(m=4, l_1=-1, l_2=4\).  Each BarHMF particle now interacts with two
shapes depending on its action values \((I_r, I_\phi)\): a dumbbell
shaped quadrupole density and a 4-lobed octopole density (e.g. two
pyramids joined at their bases or two orthogonally-oriented
dumbbells).  Each additional resonant term adds a new degree of
freedom corresponding to the slow angle action from equation
(\ref{eq:F2}).  In addition, combinations with multiple \(m\) orders
will have separate sets of coefficients for each order from equation
(\ref{eq:coefj}).

With one term, the BarHMF simulation is an ensemble of interacting
pendula.  This system is intrinsically irregular.  Indeed some of the
simulations described in Appendix \ref{sec:hmftests} show evidence of
chaotic behavior.  As we add more terms, we introduce the possibility
additional chaotic channels such as resonance overlap
\citep{Chirikov:1979} and weak chaotic transport \citep{Manos:2011}.
The BarHMF model is ideal model for investigating the details of
multiple resonance phenomena in barred systems.  This ability to study
the interactions of specific resonances is ideal for isolating and
determining their relevance to barred systems in traditional
simulations and Nature.  With only several specific resonant terms,
the computational effort will remain modest.

The examples in Appendix \ref{sec:hmftests} assume a spherical
background potential for expediency and the same two-dimensional basis
described in Appendix \ref{sec:2ddisk} and used for the simulations
Section \ref{sec:results}.  We may couple the in-plane dynamics of the
BarHMF model to other galactic components using the potential theory
described in Appendix \ref{sec:2ddisk}.  For example, we can study the
evolution of the bar in the presence of dynamical friction and
investigate the dependence on the dark-matter halo properties. The
BarHMF model presents a new opportunity to understand the changes and
redistribution of the bar orbits under torque, a subject which remains
poorly understood dynamically.  This exploration might be especially
productive with BarHMF which provides the opportunity to explicitly
identify interactions between multiple resonances.

\end{appendix}

\end{document}